\documentclass{elsart3p}

\usepackage{amsmath,amssymb,graphicx}
\usepackage{natbib}

\newcommand{\D}{{\mathrm{d}}}

\begin{document}
\begin{frontmatter}

\title{{\small Chemical Engineering Science 65 (2010) 2310-2324} \\ \vspace{10mm} Asymptotology of Chemical Reaction Networks}
\author{A.~N.~Gorban \corauthref{cor1}}
 \ead{ag153@le.ac.uk}
\address{University of Leicester, UK}
\corauth[cor1]{Corresponding author: University of Leicester, LE1
7RH, UK}
\author{O. Radulescu}
 \ead{ovidiu.radulescu@univ-rennes1.fr}
\address{IRMAR, UMR 6625, University of Rennes 1, Campus de Beaulieu, 35042 Rennes, France}
\author{A. Y. Zinovyev}
 \ead{andrei.zinovyev@curie.fr}
\address{Institut Curie, U900 INSERM/Curie/Mines ParisTech, 26 rue d'Ulm, F75248, Paris, France}

\date{}

\maketitle

\begin{abstract}
The concept of the limiting step is extended to the asymptotology
of multiscale reaction networks. Complete theory for linear
networks with well separated reaction rate constants is developed.
We present algorithms for explicit approximations of eigenvalues
and eigenvectors of kinetic matrix. Accuracy of estimates is
proven. Performance of the algorithms is demonstrated on simple
examples. Application of algorithms to nonlinear systems is
discussed.
\end{abstract}
\end{frontmatter}

\section{Introduction \label{sec1}}

Most of mathematical models that really work are simplifications
of the basic theoretical models and use in the backgrounds an
assumption that some terms are big, and some other terms are small
enough to neglect or almost neglect them. The closer consideration
shows that such a simple separation on ``small" and ``big" terms
should be used with precautions, and special culture was
developed. The name ``asymptotology" for this direction of science
was proposed by \cite{Kruskal}, but fundamental research in this
direction are much older, and many fundamental approaches were
developed by I. Newton (Newton polyhedron, and many other things).

Following \cite{Kruskal}, {\it asymptotology} is ``the art of
describing the behavior of a specified solution (or family of
solutions) of a system in a limiting case. ... The art of
asymptotology lies partly in choosing fruitful limiting cases to
examine ... The scientific element in asymptotology resides in the
nonarbitrariness of the asymptotic behavior and of its description,
once the limiting case has been decided upon."

Asymptotic behavior of rational functions of several positive
variables $k_i>0$ gives us a toy-example. Let $$R(k_1,\ldots
k_n)=P(k_1,\ldots k_n)/Q(k_1,\ldots k_n)$$ be such a function and
$P,Q$ be polynomials. To derive fruitful limiting cases we consider
logarithmic straight lines $\ln k_i=\theta_{i} \xi$  and study
asymptotical behavior of $R$ for  $\xi \to \infty$. In this
asymptotics, for almost every vector $(\theta_i)$ (outside several
hyperplanes) there exists such a {\it dominant monomial}
$R_{\infty}(k)=A \prod_i k_i^{\alpha _i}$ that $R=R_{\infty}
+o(R_{\infty})$. The function that associates a monomial with vector
$(\theta_i)$ is piecewise constant: it is constant inside some
polyhedral cones.

Implicit functions given by equations which depend on parameters
provide plenty of more interesting examples, especially in the case
when the implicit function theorem is not applicable. Some
analytical examples are presented by \cite{AndrianovManevitch2002}
and \cite{White}. Introduction of algebraic backgrounds and special
software is provided by \cite{orderPfi}.

For a difficult problem, analysis of eigenvalues and eigenvectors of
non-symmetric matrices, \cite{ViLju} studied asymptotic behavior of
spectra and spectral projectors along the logarithmic straight lines
in the space of matrices. This analysis was continued by \cite{Lid}.

We study networks of linear reactions. For a linear system with
reaction rate constants $k_i$ all the dynamical information is
contained in eigenvalues and eigenvectors of the kinetic matrix or,
more precisely, in its transformation to the Jordan normal form. It
is computationally expensive task to find this transformation for a
non-symmetric matrix which is usually stiff (\cite{Golub}).
Moreover, the answer could be very sensitive to the errors in
constants $k_i$. Nevertheless, it appears that stiffness can help us
to find a robust approximation, and in the limit when all constants
are very different (well-separated constants) the asymptotical
behavior of eigenvalues and eigenvectors follow simple explicit
expressions. Analysis of this asymptotics is our main goal.

In our approach, we study asymptotic behavior of  eigenvalues and
eigenvectors of kinetic matrices along logarithmic straight lines,
$\ln k_i=\theta_{i} \xi$ in the space of constants. We significantly
use the graph representation of chemical reaction networks and
demonstrate, that for almost every vector $(\theta_i)$ there exists
a simple reaction network which describes the dominant term of this
asymptotic. Following the asymptotology terminology (\cite{White}),
we call this simple network the {\it dominant system}. For these
dominant system there are explicit formulas for eigenvalues and
eigenvectors. The topology of dominant systems is rather simple:
they are acyclic networks without branching. This allows us to
construct the explicit asymptotics of eigenvectors and eigenvalues.
All algorithms are represented topologically by transformation of
the graph of reaction (labeled by reaction rate constants). The
reaction rate constants for dominant systems may not coincide with
constant of original network. In general, they are monomials of the
original constants.

This result fully supports the observation by \cite{Kruskal}: ``And
the answer quite generally has the form of a new system (well posed
problem) for the solution to satisfy, although this is sometimes
obscured because the new system is so easily solved that one is led
directly to the solution without noticing the intermediate step."

The dominant systems can be used for direct computation of steady
states and relaxation dynamics, especially when kinetic information
is incomplete, for design of experiments and mining of experimental
data, and could serve as a robust first approximation in
perturbation theory or for preconditioning. They can be used to
answer an important question: given a network model, which are its
critical parameters? Many of the parameters of the initial model are
no longer present in the dominant system: these parameters are
non-critical. Parameters of dominant subsystems indicate putative
targets to change the behavior of the large network.

Most of reaction networks are nonlinear, it is nevertheless useful
to have an efficient algorithm for solving linear problems. First,
nonlinear systems often include linear subsystems, containing
reactions that are (pseudo)monomolecular with respect to species
internal to the subsystem (at most one internal species is reactant
and at most one is product). Second, for binary reactions $A + B \to
...$, if concentrations of species $A$ and $B$ ($c_A, c_B$) are well
separated, say $c_A \gg c_B$ then we can consider this reaction as
$B \to ...$ with rate constant proportional to $c_A$ which is
practically constant, because its relative changes are small in
comparison to relative changes of $c_B$. We can assume that this
condition is satisfied for all but a small fraction of genuinely
nonlinear reactions (the set of nonlinear reactions changes in time
but remains small). Under such an assumption,  nonlinear behavior
can be approximated as a sequence of such systems, followed one each
other in a sequence of ``phase transitions". In these transitions,
the order relation between some of species concentrations changes.
Some applications of this approach to systems biology are presented
by \cite{RadGorZinLil2008}. The idea of controllable linearization
``by excess" of some reagents is in the background of the efficient
experimental technique of Temporal Analysis of Products (TAP), which
allows to decipher detailed mechanisms of catalytic reactions
(\cite{TAP}).

In chemical kinetics various fundamental ideas about asymptotical
analysis were developed (\cite{Klonowski1983}): quasieqiulibrium
asymptotic (QE), quasi steady-state asymptotic (QSS), lumping, and
the idea of limiting step.

Most of the works on nonequilibrium thermodynamics deal with the QE
approximations and corrections to them, or with applications of
these approximations (with or without corrections). There are two
basic formulation of the QE approximation: the thermodynamic
approach, based on entropy maximum, or the kinetic formulation,
based on selection of fast reversible reactions. The very first use
of the entropy maximization dates back to the classical work of
\cite{Gibb}, but it was first claimed for a principle of
informational statistical thermodynamics by \cite{Janes1}. A very
general discussion of the maximum entropy principle with
applications to dissipative kinetics is given in the review by
\cite{Bal}. Corrections of QE approximation with applications to
physical and chemical kinetics were developed by
\cite{GKIOeNONNEWT2001,GorKar}.

QSS was proposed by \cite{Bodenstein1913} and was elaborated into an
important tool for analysis of chemical reaction mechanism and
kinetics (\cite{Semenov1939,Christiansen1953,Helfferich1989}). The
classical QSS is based on the relative smallness of concentrations
of some of ``active" reagents (radicals, substrate-enzyme complexes
or active components on the catalyst surface)
(\cite{Aris1965,Segel89}).

Lumping analysis aims to combine reagents into ``quasicomponents"
for dimension reduction
(\cite{LumpWei1,LumpWei2,LumpLiRab1,LumpLiRab2}.

The concept of limiting step gives the limit simplification: the
whole network behaves as a single step. This is the most popular
approach for model simplification in chemical kinetics and in many
areas beyond kinetics. In the form of a {\it bottleneck} approach
this approximation is very popular from traffic management to
computer programming and communication networks. The proposed
asymptotic analysis can be considered as a wide extension of the
classical idea of limiting step (\cite{GorbaRadul2008}).

The structure of the paper is as follows. In Sec.~\ref{sec2} we
introduce basic notions and notations. We consider thermodynamic
restrictions on the reaction rate constants and demonstrate how
appear systems with arbitrary constants (as subsystems of more
detailed models). For linear networks, the main theorems which
connect ergodic properties with topology of network, are reminded.
Four basic ideas of model reduction in chemical kinetics are
described: QE, QSS, lumping analysis and limiting steps.

In Sec.~\ref{sec3}, we introduce the dominant system for a simple
irreversible catalytic cycle with limiting step. This is just a
chain of reactions which appears after deletion the limiting step
from the cycle. Even for such simple examples several new
observation are presented:
\begin{itemize}
\item{The relaxation time for a cycle with limiting step is
inverse second reaction rate constant;}
\item{For chains of reactions with well separated rate constants
left eigenvectors have coordinates close to 0 or 1, and right
eigenvectors have coordinates close to 0 or $\pm1$.}
\end{itemize}
For general reaction networks instead of linear chains appear
general acyclic non-branching networks. For them we also provide
explicit formulas for eigenvectors and their 0, $\pm1$ asymptotics
for well-separated constants (Sec.~\ref{sec4}). In
(Sec.~\ref{sec5}) the main algorithm is presented. Sec.~\ref{sec6}
is devoted to a simple demonstration of the algorithm application.
In Sec.~\ref{sec7}, we briefly discuss further corrections to
dominant systems. The estimates of accuracy are given in Appendix.

\section{Main Asymptotic Ideas in Chemical Kinetics \label{sec2}}

\subsection{Chemical Reaction Networks}

To define a chemical reaction network, we have to introduce:
\begin{itemize}
\item{a list of components (species);}
\item{a list of elementary reactions;}
\item{a kinetic law of elementary reactions.}
\end{itemize}
The list of components is just a list of symbols (labels) $A_1,...
A_n$. Each elementary reaction is represented by its {\it
stoichiometric equation}
\begin{equation}
\sum_{i} \alpha_{si} A_i \to \sum_{si} \beta_{si} A_i ,
\end{equation}
where $s$ enumerates the elementary reaction, and the non-negative
integers $\alpha_{si}$, $\beta_{si}$ are the {\it stoichiometric
coefficients}.  A stoichiomentric vector $\gamma_s$ with
coordinates $\gamma_{si}=\beta_{si}-\alpha_{si}$ is associated
with each elementary reaction.

For analysis of closed chemical systems with detailed balance it
is usual practice to group reactions in pairs, direct and inverse
reactions together, but in more general settings this is not
convenient.

A non-negative real {\it extensive} variable $N_i \geq 0$, amount of
$A_i$, is associated with each component $A_i$. It measures ``the
number of particles of that species" (in particles, or in moles).
The concentration of $A_i$ is an {\it intensive} variable: $c_i =
N_i/V$, where $V$ is volume. It is necessary to stress, that in many
practically important cases the extensive variable $V$ is neither
constant, nor the same for all components $A_i$. For more details
see, for example the book of \cite{Yab}. For simplicity, we will
consider systems with one constant volume and under constant
temperature, but it is necessary always keep in mind the possibility
to return to general equations. For that conditions, the kinetic
equations have the following form
\begin{equation}\label{ConcKinur}
\frac{\D c}{\D t}=\sum_s w_s(c,T)\gamma_s + \upsilon,
\end{equation}
where $ \upsilon$ is the vector of external fluxes normalized to
unit volume. It may be useful to represent external fluxes as
elementary reactions by introduction of new component
$\varnothing$ together with incoming and outgoing reactions
$\varnothing \to A_i$ and $A_i \to \varnothing$.

The most popular {\it kinetic law} of elementary reactions is the
{\it mass action law} for perfect systems:
\begin{equation}\label{MAL}
w_s(c,T)=k_s(T)\prod c_i^{\alpha_{si}},
\end{equation}
where ``kinetic constant" $k_s(T)$ depends on temperature $T$.
More general kinetic law, which can be used for most of non-ideal
(non-perfect) systems is
\begin{equation}\label{MDD}
w_s(c,T)=\varphi_s\exp\left(\frac{1}{RT}\sum_i \alpha_{si} \mu_i
\right),
\end{equation}
where $R$ is the universal gas constant, $\mu_i$ is the chemical
potential, $\mu_i=\frac{\partial F(N,T,V)}{\partial
N_i}=\frac{\partial G(N,T,P)}{\partial N_i}$, $F$ is the Helmgoltz
free energy, $G$ is the Gibbs energy (free enthalpy), $P$ is
pressure and $\varphi_s >0$ is an intensive variable, kinetic
factor, which can depend on any set of intensive variables, first
of all, on $T$.

Chemical thermodynamics (\cite{PrigogineDefay1954}) provides tools
of choice for stability analysis of reaction networks
(\cite{ProcacciaRoss}) and chemical reactors (\cite{Aris1965}). The
laws of thermodynamics have been used for analyzing of structural
stability of process systems by \cite{Hangos2004}. In general
reaction network coefficients $k_s$ (\ref{MAL}) or $\varphi_s$
(\ref{MDD}) are not independent.  In order to respect the second law
of thermodynamics, they should satisfy some equations and
inequalities. The most famous sufficient condition gives the {\it
principle of detailed balance}. Let us group the elementary
reactions in pairs, direct and inverse reactions, and mark the
variables for direct reactions by superscript $+$, and for inverse
reactions by $-$. Then the principle of detailed balance for general
kinetics (\ref{MDD}) reads:
\begin{equation}\label{DetBalMDD}
\varphi_s^+ =\varphi_s^-
\end{equation}
(\cite{Feinberg1972}). For the isothermal  mass action law the
principle of detailed balance can be formulated as follows: there
exists a strictly positive point $c^*$ of detailed balance, at
this point
\begin{equation}\label{DetBalMAL}
w_s^+(c^*)=w_s^-(c^*)
\end{equation}
for all $s$. This is, essentially, the same principle: if we
substitute in the general reaction rate (\ref{MDD}) the fraction
$\mu_i/RT$ by $\ln(c_i/c_i^*)$, then we will get the mass action
law, and $\varphi_s^+ =\varphi_s^-$. The principle of detailed
balance is closely related to the microreversibility and Onsager
relations.

More general condition was invented by \cite{Stueck} for the
Boltzmann equation. He produced them from the $S$-matrix unitarity
(the quantum complete probability formula). For the general law
(\ref{MDD}) without direct-inverse reactions grouping for any
state the following identity holds:
\begin{equation}\label{Stuck}
\begin{split}
&\sum_s \varphi_s \exp\left(\frac{1}{RT}\sum_i \alpha_{si} \mu_i
\right)\\ &\equiv \sum_s \varphi_s \exp\left(\frac{1}{RT}\sum_i
\beta_{si} \mu_i \right).
\end{split}
\end{equation}
Even more general condition which guarantees the second law and
has clear microscopic sense (the complete probability does not
increase) was obtained by \cite{Gorban1984}: for any state
\begin{equation}\label{SemiStuck}
\begin{split}
&\sum_s \varphi_s \exp\left(\frac{1}{RT}\sum_i \alpha_{si} \mu_i
\right)\\ &\geq \sum_s \varphi_s \exp\left(\frac{1}{RT}\sum_i
\beta_{si} \mu_i \right).
\end{split}
\end{equation}
To obtain formulas for the isothermal mass action law, it is
sufficient just to apply the general law (\ref{MDD}) with constant
$\varphi_s$ to the perfect free energy $F=RT \sum_i c_i (\ln c_i +
\mu_{i0})$ with constant $\mu_{i0}$. More detailed analysis  was
presented, by \cite{Gorban1984}.

In any case, reaction constants are dependent, and this dependence
guarantees stability of equilibrium and existence of global
thermodynamic Lyapunov functions for closed systems
(\ref{ConcKinur}) with $\upsilon =0$. Nevertheless, we often study
equations for such systems with oscillations, bifurcations, chaos,
and other effects, which are impossible in systems with global
Lyapunov function. Usually this means that we study a subsystem of
a large system, where some of concentrations do not change because
they are stabilized by external fluxes or by a large external
reservoir. These constant (or very slow) concentrations are
included into new reaction constants, and after this redefinition
they can loose any thermodynamic property.

\subsection{Linear Networks and Ergodicity}

In this Sec., we consider a general network of linear reactions.
This network is represented as a directed graph (digraph)
(\cite{Temkin1996}): vertices correspond to components $A_i$, edges
correspond to reactions $A_i \to A_j$ with kinetic constants $k_{ji}
> 0$. For each vertex, $A_i$, a positive real variable $c_i$
(concentration) is defined. A basis vector $e^i$ corresponds to
$A_i$ with components $e^i_j=\delta_{ij}$, where $\delta_{ij}$ is
the Kronecker delta. The kinetic equation for the system is
\begin{equation}\label{kinur}
\frac{\D c_i}{\D t}=\sum_j (k_{ij} c_j - k_{ji}c_i),
\end{equation}
or in vector form: $\dot{c} = Kc$. We don't assume any special
relation between constants, and consider them as independent
quantities. The thermodynamic restrictions on constants are not
applicable here because, in general, we study pseudomonomolecular
systems which are subsystems of larger nonlinear systems and don't
represent by themselves closed monomolecular systems.

For any  network of linear reactions the matrix of kinetic
coefficients $K$ has  the following properties:
\begin{itemize}
\item{non-diagonal elements of $K$ are non-negative;}
\item{diagonal elements of $K$ are non-positive;}
\item{elements in each column of $K$ have zero sum.}
\end{itemize}
For any $K$ with these properties there exists a network of linear
reactions with kinetic equation $\dot{c}=Kc$. This family of
matrices coincide with the family of generators of finite Markov
chains, and this class of kinetic equations coincide with the class
of inverse Kolmogorov's equations or master equations for the finite
Markov chains in continuous time (\cite{MeynMarkCh2009};
\cite{MeynNets2007}).

A {\it linear conservation law} is a linear function defined on the
concentrations $b(c)= \sum_{i} b_i c_i$, whose value is preserved by
the dynamics \eqref{kinur}. The conservation laws coefficient
vectors $b_i$ are left eigenvectors of the matrix $K$ corresponding
to the zero eigenvalue. The set of all the conservation laws forms
the left kernel of the matrix $K$. Equation \eqref{kinur} always has
a linear conservation law: $b^0(c)=\sum_i c_i = {\rm const}$. If
there is no other independent linear conservation law, then the
system is {\it weakly ergodic}.

A set $E$ is {\it positively invariant} with respect to kinetic
equations (\ref{kinur}), if any solution $c(t)$ that starts in $E$
at time $t_0$ ($c(t_0) \in E$) belongs to $E$ for $t>t_0$ ($c(t) \in
E$ if $t>t_0$). It is straightforward to check that the standard
simplex $\Sigma = \{c\,| \, c_i \geq 0, \, \sum_i c_i =1\}$ is
positively invariant set for kinetic equation (\ref{kinur}): just to
check that if  $c_i=0$ for some $i$, and all $c_j \geq 0$ then
$\dot{c}_i \geq 0$. This simple fact immediately implies the
following properties of ${K}$:
\begin{itemize}
\item{All eigenvalues $\lambda$ of ${K}$ have non-positive real
parts, $Re \lambda \leq 0$, because solutions cannot leave
$\Sigma$ in positive time;}
\item{If $Re \lambda = 0$ then $\lambda = 0$, because intersection of
$\Sigma$ with any plane is a polygon, and a polygon cannot be
invariant with respect to rotations to sufficiently small angles;}
\item{The Jordan cell of ${K}$ that corresponds to zero
    eigenvalue is diagonal -- because all solutions should
    be bounded in $\Sigma$ for positive time.}
\item{The shift in time operator $\exp({K} t)$ is a
    contraction in the $l_1$ norm for $t>0$.}
\item{For weakly ergodic systems there exists such a
    monotonically decreasing function $\delta(t)$ ($t>0$,
    $0<\delta(t)<1$, $\delta(t) \to 0$ when $t\to \infty$)
    that for any two solutions of (\ref{kinur}) $c(t),
    c'(t) \in \Sigma$
\begin{equation}
\sum_i |c_i(t) -c'_i(t)| \leq \delta(t) \sum_i |c_i(0) -c'_i(0)| \ .
\end{equation} }
\end{itemize}

The {\it ergodicity coefficient} $\delta(t)$ was introduced by
\cite{DobrushinErgCoeff1956} (see also  a book by
\cite{Seneta1981}). It can be estimated using the structure of the
network graph (\cite{Ocherki,MeynNets2007}).

Two vertices are called adjacent if they share a common edge. A path
is a sequence of adjacent vertices. A graph is connected if any two
of its vertices are linked by a path. A maximal connected subgraph
of graph $G$ is called a connected component of $G$. Every graph can
be decomposed into connected components.

A directed path is a sequence of adjacent edges where each step
goes in direction of an edge. A vertex $A$ is {\it reachable} from
a vertex $B$, if there exists a directed path from $B$ to $A$.

A nonempty set $V$ of graph vertices forms a {\it sink}, if there
are no directed edges from $A_i \in V$ to any $A_j \notin V$. For
example, in the reaction graph $A_1\leftarrow A_2 \rightarrow A_3$
the one-vertex sets $\{A_1\}$ and $\{A_3\}$ are sinks. A sink is
minimal if it does not contain a strictly smaller sink. In the
previous example, $\{A_1\}$, $\{A_3\}$  are minimal sinks. Minimal
sinks are also called ergodic components.

A digraph is strongly connected, if every  vertex $A$ is reachable
from any other vertex $B$. Ergodic components are maximal strongly
connected subgraphs of the graph, but inverse is not true: there
may exist maximal strongly connected subgraphs that have outgoing
edges and, therefore, are not sinks.

The weak ergodicity of the network follows from its topological
properties.

{\bf Theorem 1.} The following properties are equivalent (and each
one of them can be used as an alternative definition of weak
ergodicity):
\begin{enumerate}
\item{There exist the only independent linear conservation law for
kinetic equations (\ref{kinur}) (this is $b^0(c)=\sum_i c_i = {\rm
const}$).}
 \item{For any normalized initial state $c(0)$ ($b^0(c)=1$)
there exists a limit state $$c^*= \lim_{t\rightarrow \infty }
\exp(Kt) \, c(0)$$ that is the same for all normalized initial
conditions: For all $c$, $$\lim_{t\rightarrow \infty } \exp(Kt) \,
c = b^0(c)  c^*.$$}
\item{For each two vertices  $A_i, \: A_j \: (i
\neq j)$ we can find such a vertex $A_k$ that is reachable both
from $A_i$ and from $A_j$. This means that the following structure
exists:
 $$A_i \to \ldots \to A_k \leftarrow \ldots \leftarrow A_j.$$
One of the paths can be degenerated: it may be $i=k$ or $j=k$.}
\item{The network has only one minimal sink (one ergodic
component).$\square$}
\end{enumerate}
The proof of this theorem could be extracted from detailed books
about Markov chains and networks
(\cite{MeynNets2007,VanMieghem2006}). In its present form it was
published by \cite{Ocherki} with explicit estimations of ergodicity
coefficients.

For every monomolecular kinetic system, the maximal number of
independent linear conservation laws (i.e. the geometric
multiplicity of the zero eigenvalue of the matrix $K$) is equal to
the maximal number of disjoint ergodic components (minimal sinks).

\subsection{Quasi-equilibrium (QE) or Fast Equilibrium}

Quasi-equilibrium approximation uses the assumption that a group of
reactions is much faster than other and goes fast to its
equilibrium. We use below superscripts `$^{\rm f}$' and `$^{\rm s}$'
to distinguish fast and slow reactions. A small parameter appears in
the following form
\begin{equation}\label{ConcKinurQE}
\begin{split}
\frac{\D c}{\D t}= &\sum_{\sigma, \ {\rm slow}} w_{\sigma}^{\rm
s}(c,T)\gamma_{\sigma}^{\rm s} + \frac{1}{\varepsilon}
\sum_{\varsigma, \ {\rm fast}} w^{\rm
f}_{\varsigma}(c,T)\gamma_{\varsigma}^{\rm f},
\end{split}
\end{equation}
To separate variables, we have to study the spaces of linear
conservation law of the initial system (\ref{ConcKinurQE}) and of
the fast subsystem
 $$\frac{\D c}{\D t}=\frac{1}{\varepsilon} \sum_{\varsigma, \ {\rm
 fast}} w^{\rm f}_{\varsigma}(c,T)\gamma_{\varsigma}^{\rm f}$$
If they coincide, then the fast subsystem just dominates, and there
is no fast-slow separation for variables (all variables are either
fast, or constant). But if there exist additional linearly
independent linear conservation laws for the fast system, then let
us introduce new variables: linear functions $b^1(c),... b^n(c)$,
where $b^1(c),... b^m(c)$ is the basis of the linear conservation
laws for the initial system, and $b^{1}(c),... b^{m+l}(c)$ is the
basis of the linear conservation laws for the fast subsystem. Then
$b^{m+l+1}(c),... b^n (c)$ are fast variables, $b^{m+1}(c),...
b^{m+l}(c)$ are slow variables, and $b^{1}(c),... b^{m}(c)$ are
constant. The {\it quasi-equilibrium manifold} is given by the
equations $\sum_{\varsigma} w^{\rm
f}_{\varsigma}(c,T)\gamma_{\varsigma}^{\rm f}=0$ and for small
$\varepsilon$ it serves as an approximation to a slow manifold. In
the old and standard approach it is assumed that system
(\ref{ConcKinurQE}) as well as system of fast reactions satisfies
the thermodynamic restrictions, and the quasi-equilibrium is just a
{\it partial thermodynamic equilibrium}, and could be defined by
conditional extremum of thermodynamic functions. This guarantees
global stability of fast subsystems and all the classical singular
perturbation theory like Tikhonov theorem could be applied.

Recently, \cite{Daoutidis} took notice that this type of reasoning
does not require classical thermodynamic restrictions on
constants. For example, let us consider the mass action law
kinetics and group the reactions in pairs, direct and inverse
reactions. If the set of stoichiometric vectors for fast reactions
is linearly independent, then for this system the detailed balance
principle holds (obviously), and it demonstrates the
``thermodynamic behaviour" without connection to classical
thermodynamics. This case of ``stoichiometrically independent fast
reactions" can be generalized for irreversible reactions too
(\cite{Daoutidis}). For such fast system the quasiequilbrium
manifold has the same nice properties as for thermodynamic partial
equilibrium, and approximates slow dynamics for sufficiently small
$\varepsilon$.

There are other classes of mass action law subsystems with such a
``quasi-thermodynamic" behaviour, which depends on structure, but
not on constants. For example, any system of reactions without
interactions has such a property (\cite{without}). These reactions
have the form $\alpha A_i \to \sum ...$: any linear reaction are
allowed, as well as reactions like $2A_i \to A_j+A_k$, $3 A_i \to
A_j+ A_k + A_l$, etc. All such fast subsystems can serve for
quasi-equilibrium approximation, because for them dynamics is
globally stable.

Quasi-equilibrium manifold approximates exponentially attractive
slow manifold and is used in many areas of kinetics either as
initial approximation for slow motion, or just by itself (more
discussion and further references are presented by \cite{GorKar}).

\subsection{Quasi Steady-State (QSS) or Fast Species}

The quasi steady-state (or pseudo steady state) assumption was
invented in chemistry for description of systems with radicals or
catalysts. In the most usual version the species are split in two
groups with concentration vectors $c^{\rm s}$ (``slow" or basic
components) and $c^{\rm f}$ (``fast intermediates"). For catalytic
reactions there is additional balance for $c^{\rm f}$, amount of
catalyst, usually it is just a sum $b_{\rm f}=\sum_i c^{\rm f}_i$.
The amount of the fast intermediates is assumed much smaller than
the amount of the basic components, but the reaction rates are of
the same order, or even the same (both intermediates and slow
components participate in the same reactions). This is the source
of a small parameter in the system. Let us scale the
concentrations $c^{\rm f}$ and $c^{\rm s}$ to the compatible
amounts. After that, the fast and slow time appear and we could
write $\dot{c}^{\rm s} = W^{\rm s}(c^{\rm s},c^{\rm f})$,
$\dot{c}^{\rm f} =\frac{1}{\varepsilon} W^{\rm f}(c^{\rm s},c^{\rm
f})$, where $\varepsilon$ is small parameter,  and functions
$W^{\rm s},W^{\rm f}$ are bounded and have bounded derivatives
(are ``of the same order"). We can apply the standard singular
perturbation techniques. If dynamics of fast components under
given values of slow concentrations is stable, then the slow
attractive manifold exists, and its zero approximation is given by
the system of equations $W^{\rm f}(c^{\rm s},c^{\rm f})=0$.
Bifurcations in fast system correspond to critical effects,
including ignition and explosion.

This scheme was analyzed many times with plenty of details,
examples, and some complications. Exhaustive case study of the
simplest enzyme reaction was provided by \cite{Segel89} . For
heterogenious catalytic reactions, the book by \cite{Yab} gives
analysis of scaling of fast intermediates (there are many kinds of
possible scaling). In the context of the Computational Singular
Perturbation (CSP) approach, \cite{Lam1993} and \cite{LamGous1994}
developed concept of the CSP radicals. \cite{InChLANL,GorKar}
considered QSS as initial approximation for slow invariant manifold.
Analysis of the error of the QSS was provided by \cite{Tomlin1993}.

The QE approximation is also extremely popular and useful. It has
simpler dynamical properties (respects thermodynamics, for example,
and gives no critical effects in fast subsystems of closed systems).
Nevertheless, neither radicals in combustion, nor intermediates in
catalytic kinetics are, in general, close to quasi-equilibrium. They
are just present in much smaller amount, and when this amount grows,
then the QSS approximation fails.

The simplest demonstration of these two approximation gives the
simple reaction: $S+E\leftrightarrow SE \to P+E$ with reaction rate
constants $k^{\pm}_1$ and $k_2$. The only possible quasi-equilibrium
appears when the first equilibrium is fast: $k^{\pm}_1 =
\kappa^{\pm}/\varepsilon$. The corresponding slow variable is
$C^s=c_S+c_{SE}$, $b_E=c_E+c_{SE}=const$. For the QE manifold we get
a quadratic equation $\frac{k_1^-}{k_1^+}c_{SE}=c_Sc_E=
(C^s-c_{SE})(b_E-c_{SE})$. This equation gives the explicit
dependence $c_{SE}(C^s)$, and the slow equation reads
$\dot{C}^s=-k_2 c_{SE}(C^s)$, $C^s+c_P=b_S=const$.

For the QSS approximation of this reaction kinetics, under
assumption $b_E \ll b_S$, we have fast intermediates $E$ and $SE$.
For the QSS manifold there is a linear equation
$k^+_1c_Sc_E-k_1^-c_{SE} - k_2c_{SE}=0$, which gives us the
explicit expression for $c_{SE}(c_S)$: $c_{SE}=k_1^+ c_S
b_E/(k_1^+ c_S+k_1^-+k_2)$ (the standard Michaelis--Menten
formula). The slow kinetics reads $\dot{c}_S=-k_1^+c_S
(b_E-c_{SE}(c_S)) + k_1^-c_{SE}(c_S)$. The difference between the
QSS and the QE in this example is obvious.

The terminology is not rigorous, and often QSS is used for all
singular perturbed systems, and QE is applied only for the
thermodynamic exclusion of fast variables by the maximum entropy (or
minimum of free energy, or extremum of another relevant
thermodynamic function) principle (MaxEnt). This terminological
convention may be convenient. Nevertheless, without any relation to
terminology, the difference between these two types of introduction
of a small parameter is huge. There exists plenty of generalizations
of these approaches, which aim to construct a slow and (almost)
invariant manifold, and to approximate fast motion as well. The
following references can give a first impression about these
methods: Method of Invariant Manifolds (MIM)
(\cite{Roussel91,GorKar}, Method of Invariant Grids (MIG), a
discrete analogue of invariant manifolds (\cite{Grids}),
Computational Singular Perturbations (CSP)
(\cite{Lam1993,LamGous1994,ZaKapers}) Intrinsic Low-Dimensional
Manifolds (ILDM)  by \cite{Maas}, developed further in series of
works by \cite{BGGMaas2006}), methods based on the Lyapunov
auxiliary theorem (\cite{KazKraLya}).

\subsection{Lumping Analysis}

\cite{Wei62} demonstrated that for (pseudo)monomolecular systems
there exist linear combinations of concentrations which evolve in
time independently. These linear combinations (quasicomponents)
correspond to the left eigenvectors of kinetic matrix: if $l K=
\lambda l$ then $\D (l,c)/\D t= (l,c) \lambda $, where the standard
inner product $(l,c)$ is concentration of a quasicomponent. They
also demonstrated how to find these quasicomponents in a properly
organized experiment.

This observation gave rise to a question: how to lump components
into proper quasicomponents to guarantee the autonomous dynamics
of the quasicomponents with appropriate accuracy. Wei and Kuo
studied conditions for exact (\cite{LumpWei1}) and approximate
(\cite{LumpWei2}) lumping in monomolecular and pseudomonomolecular
systems. They demonstrated that under certain conditions large
monomolecular system could be well--modelled by lower--order
system.

More recently, sensitivity analysis and Lie group approach were
applied to lumping analysis (\cite{LumpLiRab1,LumpLiRab2}), and
more general nonlinear forms of lumped concentrations are used
(for example, concentration of quasicomponents could be rational
function of $c$).

\cite{LumpParal1stOrd} studied lumping-analysis of mixtures with
many parallel first order reactions. \cite{LumpingStructure}
generalized these results and characterized those lumping schemes
which preserve the kinetic structure of the original system.
\cite{LumpingOBservability} placed lumping analysis in the linear
systems theory and demonstrated the relationships between
lumpability and the concepts  of observability, controllability
and minimal realization. \cite{NonstiffAtmospheric2002} considered
the lumping procedures as efficient techniques leading to nonstiff
systems and demonstrated efficiency of developed algorithm on
kinetic models of atmospheric chemistry. \cite{OptimalLumping2008}
formulated an optimal lumping problem as a mixed integer nonlinear
programming (MINLP) and demonstrated that it can be efficiently
solved with a stochastic optimization method, Tabu Search (TS)
algorithm.

The power of lumping using a time-scale based approach was
demonstrated by \cite{LumpAthmTime}. This computationally cheap
approach combines ideas of sensitivity analysis with simple and
useful grouping of species with similar lifetimes and similar
topological properties caused by connections of the species in the
reaction networks. The lumped concentrations in this approach are
simply sums of concentrations in groups. For example, species with
similar composition and functionalities could be lumped into one
single representative species (\cite{LumpingCombust2008}).

Lumping analysis based both on mathematical arguments and
fundamental physical and chemical properties of the components is
now one of the main tools for model reduction in highly
multicomponent systems, such as  the hydrocarbon mixture  in
petroleum chemistry (\cite{LumpingPetroleum2004}) or biochemical
networks in systems biology (\cite{LumpingLiving2006}). The
optimal solution of lumping problem often requires the exhaustive
search, and instead of them various heuristics are used to avoid
combinatorial explosion. For the lumping analysis of the systems
biology models \cite{PropLumpSysBio2009} developed a heuristic
greedy search strategy which allowed them to avoid the exhaustive
search of proper lumped components.

Procedures of lumping analysis form a part of general algebra of
model building and model simplification transformations.
\cite{HangosProcessBook} applied formal methods of computer
science and artificial intelligence for analysis of this algebra.
In particular, a formal method for defining syntax and semantics
of process models has been proposed.

The modern systems and control theory provides efficient tools for
lumping--analysis. The so-called balanced model reduction was
invented in late 1970s (\cite{Moore1981}). For a linear system a set
of ``target variables" is selected. The dimension of the system $n$
is large, while the number of the target variables, for example,
inputs $m$ and outputs $p$, usually satisfies $m,p \ll n$.  The {\it
balanced model reduction problem} can be stated as follows
(\cite{GugercinAntoulas2004}): find a reduced order system such that
the following properties are satisfied:
\begin{enumerate}
\item{The approximation error in the target variables is small, and
there exists a global error bound.}
\item{System properties, like stability and passivity, are preserved.}
\item{The procedure is computationally efficient.}
\end{enumerate}
In large dimensions, special efforts are needed to resolve the
accuracy/efficiency dilemma and to find efficiently the approximate
solution of the model reduction problem
(\cite{AntoulasSorensen2002}).

Various methods for balanced truncation are developed: Lyapunov
balancing, stochastic balancing, bounded real balancing,  positive
real balancing, and frequency weighted balancing
(\cite{GugercinAntoulas2004}). Nonlinear generalizations are
proposed as well
(\cite{MarsdenTruncation2000,CondonTruncation2004}).

\subsection{Limiting Steps}

In the IUPAC Compendium of Chemical Terminology (2007) one can find
a definition of limiting steps. \cite{R-cont}: ``A rate-controlling
(rate-determining or rate-limiting) step in a reaction occurring by
a composite reaction sequence is an elementary reaction the rate
constant for which exerts a strong effect -- stronger than that of
any other rate constant -- on the overall rate."

Let us complement this definition by additional comment: usually
when people are talking about limiting step they expect
significantly more: there exists  a rate constant which exerts such
a strong effect on the overall rate that the effect of all other
rate constants together is significantly smaller. For the IUPAC
Compendium definition a rate-controlling step always exists, because
among the control functions generically exists the biggest one. On
the contrary,  for the notion of limiting step that is used in
practice, there exists a difference between systems with limiting
step and systems without limiting step.

During XX century, the concept of the limiting step was revised
several times. First simple idea of a ``narrow place" (the least
conductive step) could be applied without adaptation only to a
simple cycle or a chain of irreversible steps that are of the
first order (see Chap. 16 of the book \cite{Johnston} or the paper
by \cite{Boyd}). When researchers try to apply this idea in more
general situations they meet various difficulties such as:
\begin{itemize}
 \item{Some reactions have to be ``pseudomonomolecular."
 Their constants depend on
 concentrations of outer components, and are constant only
 under condition that these outer components are present in
 constant concentrations, or change sufficiently slow
 (i.e. are present in significantly bigger amount).}
 \item{Even under fixed or slow outer components concentration,
 the simple ``narrow place" behaviour
 could be spoiled by branching or by reverse reactions.
 The simplest example is given by the cycle:
 $A_1 \leftrightarrow A_2 \to A_3 \to A_1$. Even if the constant of the last step
 $A_3 \to A_1$ is the smallest one, the stationary rate may be much smaller than
 $k_3b$ (where $b$ is the overall balance of concentrations, $b=c_1+c_2+c_3$), if the
 constant of the reverse reaction $A_2 \to A_1$ is sufficiently big.}
\end{itemize}

In a series of papers, \cite{Northrop1,Northrop2} clearly
explained these difficulties and suggested that the concept of
rate--limiting step is ``outmoded". Nevertheless, the main idea of
limiting is so attractive that  Northrop's arguments stimulated
the search for modification and improvement of the main concept.

\cite{Ray} proposed the use of sensitivity analysis. He considered
cycles of reversible reactions and suggested a definition: {\it
The rate--limiting step in a reaction sequence is that forward
step for which a change of its rate constant produces the largest
effect on the overall rate}.

Ray's approach was revised by \cite{BrownCo} from the system
control analysis point of view (see the book of \cite{CorBow}).
They stress again that there is no unique rate--limiting step
specific for an enzyme, and this step, even if it exists, depends
on substrate, product and effector concentrations.

Near critical conditions the {\it critical simplification}
appears, which is also a type of limitation, because some
reactions become critically important (\cite{YabCritSimpl})

Two classical examples of limiting steps demonstrate us the chain of
linear reaction and the linear catalytic cycle, when they include a
reaction which is significantly slower, than other reactions.

A linear chain of reactions, $A_1 \to A_2 \to ... A_n$, with
reaction rate constants $k_i$ (for $A_i \to A_{i+1}$), gives the
first example of limiting steps. Let the reaction rate constant
$k_q$ be the smallest one. Then we expect the following behaviour of
the reaction chain in time scale $\gtrsim 1/k_q$: all the components
$A_1,... A_{q-1}$ transform fast into $A_q$, and all the components
$A_{q+1}, ... A_{n-1}$ transform fast into $A_n$, only two
components, $A_q$ and $A_n$ are present (concentrations of other
components are small) , and the whole dynamics in this time scale
can be represented by a single reaction $A_q \to A_n$ with reaction
rate constant $k_q$. This picture becomes more exact when $k_q$
becomes smaller with respect to other constants.

The catalytic cycle is one of the most important substructures
that we study in reaction networks. In the reduced form the
catalytic cycle is a set of linear reactions: $$A_1 \to A_2 \to
\ldots A_n \to A_1.$$ Reduced form means that in reality some of
these reaction are not monomolecular and include some other
components (not from the list $A_1, \ldots A_n$). But in the study
of the isolated  cycle dynamics, concentrations of these
components are taken as constant and are included into kinetic
constants of the cycle linear reactions.

For the constant of  elementary reaction $A_i \to $ we use the
simplified notation $k_i$ because the product of this elementary
reaction  is known, it is $A_{i+1}$ for $i<n$ and $A_1$ for $i=n$.
The elementary reaction rate is $w_i = k_i c_i$, where $c_i$ is the
concentration of $A_i$. The kinetic equation is:
\begin{equation}\label{kinCyc}
\dot{c}_i = k_{i-1} c_{i-1} - k_i c_i,
\end{equation}
where by definition $c_0=c_n$, $k_0=k_n$, and $w_0=w_n$. In the
stationary state ($\dot{c}_i = 0$), all the $w_i$ are equal:
$w_i=w$. This common rate $w$ we call the cycle stationary rate,
and
\begin{equation}\label{CycleRate}
w = \frac{b}{\frac{1}{k_1}+\ldots \frac{1}{k_n}}; \; \; c_i
=\frac{w}{k_i},
\end{equation}
where $b=\sum_i c_i$ is the conserved quantity for reactions in
constant volume. Let one of the constants, $k_{\min}$,  be much
smaller than others (let it be $k_{\min} = k_n$):
\begin{equation}\label{LimStepCycle}
k_i \gg k_{\min} \ \ {\rm if} \ \ i\neq n \ .
\end{equation}
In this case, in linear approximation $w= k_n b$,
\begin{equation}\label{CycleLimRateLin}
c_n = b\left(1 -  \sum_{i<n}\frac{k_n}{k_i}\right),\ {\rm and}  \;
c_i = b \frac{k_n}{k_i} \ {\rm for} \ i\neq n \ .
\end{equation}

The simplest zero order approximation for the steady state gives
\begin{equation}\label{CycleLimZero}
c_n = b,  \; c_i = 0\; (i \neq n).
\end{equation}
This is trivial: all the concentration is collected at the starting
point of the ``narrow place," but may be useful as an origin point
for various approximation procedures.

So, the stationary rate of a cycle is determined by the smallest
constant, $k_{\min}$, if it is much smaller than the constants of
all other reactions (\ref{LimStepCycle}):
\begin{equation}\label{limitation}
w\approx k_{\min} b .
\end{equation}
In that case we say that the cycle has a limiting step with
constant $k_{\min}$.

\section{Dynamics of Catalytic Cycle with Limiting Step \label{sec3}}
\subsection{Eigenvalues}

There is significant difference between the examples of limiting
steps for the chain of reactions and for irreversible cycle. For the
chain, the steady state does not depend on nonzero rate constants.
It is just $c_n=b, c_1=c_2=...=c_{n-1}=0$. The smallest rate
constant $k_q$ gives the smallest positive eigenvalue, the
relaxation time is $\tau = 1/k_q$. The corresponding approximation
of eigenmode (right eigenvector) $r^1$ has coordinates:
$r^1_1=...=r^1_{q-1}=0$, $r^1_q=1$, $r^1_{q+1}=...=r^1_{n-1}=0$,
$r_n=-1$. This exactly corresponds to the statement that the whole
dynamics in the time scale $\gtrsim 1/k_q$ can be represented by a
single reaction $A_q \to A_n$ with reaction rate constant $k_q$. The
left eigenvector for eigenvalue $k_q$ has approximation $l^1$ with
coordinates $l^1_1=l^1_2=...=l^1_q=1$, $l^1_{q+1}=...=l^1_n=0$. This
vector provides the almost exact {\it lumping} on time scale
$\gtrsim 1/k_q$. Let us introduce a new variable $c_{\rm
lump}=\sum_i l_i c_i$, i.e. $c_{\rm lump} =c_1+c_2+...+c_q$. For the
time scale $\gtrsim 1/k_q$ we can write $c_{\rm lump}+c_n\approx b$,
$\D c_{\rm lump} /\D t \approx -k_q c_{\rm lump}$, $\D c_n /\D t
\approx k_q c_{\rm lump}$.

In the example of a cycle, we approximate the steady state, that
is, the right eigenvector $r^0$  for zero eigenvalue (the left
eigenvector is known and corresponds to the main linear balance
$b$: $l^0_i\equiv 1$). In the zero-order approximation, this
eigenvector has coordinates $r^0_1=...=r^0_{n-1}=0$, $r^0_n=1$.

If ${k_n}/{k_i}$ is  small for all $i<n$, then the kinetic behaviour
of the cycle is determined by a linear chain of $n-1$ reactions $A_1
\to A_2 \to ... A_n$, which we obtain after cutting the limiting
step. The characteristic equation for an irreversible cycle,
$\prod_{i=1}^n (\lambda + k_i) - \prod_{i=1}^n k_i =0$, tends to the
characteristic equation for the linear chain, $\lambda
\prod_{i=1}^{n-1} (\lambda + k_i)=0$, when $k_n \to 0$.

The characteristic equation for a cycle with limiting step ($k_n/k_i
\ll 1$) has one simple zero eigenvalue that corresponds to the
conservation law $\sum c_i = b$ and $n-1$ nonzero eigenvalues
\begin{equation}\label{cycle spectra}
\lambda_i = - {k_i} + \delta_i \; (i<n).
\end{equation}
where $\delta_i \to 0$ when $\sum_{i<n}\frac{k_n}{k_i} \to 0$.

A cycle with limiting step (\ref{kinCyc}) has real eigenspectrum and
demonstrates monotonic relaxation without damped oscillations. Of
course, without limitation such oscillations could exist, for
example, when all $k_i\equiv k>0$, ($i=1,...n$).

The relaxation time of a stable linear system (\ref{kinCyc}) is, by
definition, $\tau = 1/\min \{Re (-\lambda_i)\}$ ($\lambda \neq 0$).
For small $k_n$, $\tau \approx 1/k_{\tau}$, $k_{\tau}=\min \{k_i\}$,
($i=1,... n-1$). In other words, for a cycle with limiting step,
$k_{\tau}$ is the second slowest rate constant: $k_{\min} \ll
k_{\tau}\leq ... $.

\subsection{Eigenvectors for Reaction Chain and for Catalytic Cycle with Limiting Step}

Let the irreversible cycle include a limiting step: $k_n \ll k_i$
($i=1,...,n-1$) and, in addition, $k_n \ll |k_i-k_j|$
($i,j=1,...,n-1$, $i\neq j$), then the eigenvectors of the kinetic
matrix almost coincide with the eigenvectors for the linear chain of
reactions $A_1 \to A_2 \to ... A_n$, with reaction rate constants
$k_i$ (for $A_i \to A_{i+1}$) (\cite{GorbaRadul2008}).

The kinetic equation for the linear chain is
\begin{equation}\label{chainkin}
\dot{c_i}=k_{i-1} c_{i-1}-k_i c_i,
\end{equation}
The coefficient matrix $K$ of these equations is very simple. It
has nonzero elements only on the main diagonal, and one position
below. The eigenvalues of $K$ are $-k_i$ ($i=1,...n-1$) and 0. The
left and right eigenvectors for 0 eigenvalue, $l^0$ and $r^0$,
are:
\begin{equation}\label{chain0eigen}
l^0=(1,1,...1), \;\; r^0=(0,0,...0,1),
\end{equation}
all coordinates of $l^0$ are equal to 1, the only nonzero
coordinate of $r^0$ is $r^0_n$ and we represent vector--column
$r^0$ in row.

Below we use explicit form of $K$ left and right eigenvectors. Let
vector--column $r^i$ and vector--row $l^i$ be right and left
eigenvectors of $K$ for eigenvalue $-k_i$. For coordinates of
these eigenvectors we use notation $r^i_j$ and $l^i_j$. Let us
choose a normalization condition $r^i_i=l^i_i=1$. It is
straightforward to check that $r^i_{j}=0$ $(j<i)$ and $l^i_{j}=0$
$(j>i)$, $r^i_{j+1} =k_j r_j /(k_{j+1}-k_i)$ $(j\geq i)$ and
$l^i_{j-1}=k_{j-1} l_j/(k_{j-1}-k_j)$ $(j \leq i)$, and
\begin{equation}\label{ChainEigen}
 r^i_{i+m}=\prod_{j=1}^m \frac{k_{i+j-1}}{k_{i+j}-k_i};
 \; l^i_{i-m}=\prod_{j=1}^m \frac{k_{i-j}}{k_{i-j}-k_i}.
\end{equation}
It is convenient to introduce formally $k_0=0$. Under selected
normalization condition, the inner product of eigenvectors is:
$l^i r^j = \delta_{ij}$, where $\delta_{ij}$ is the Kronecker
delta.

\begin{figure}[t]
\centering{ \includegraphics[width=70mm]{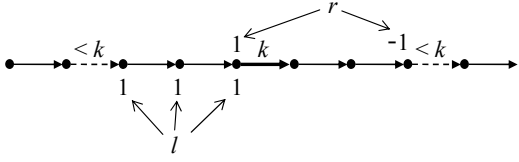} }
\caption{\label{FigChainEigen} Graphical representation of
eigenvectors approximation for the linear chain of reactions with
well separated constants. To find the left ($l$) and right ($r$)
eigenvectors for eigenvalue $k$ it is necessary to delete from the
chain all the reactions with the rate constants $<k$ (dashed
lines) and to find the maximal connected interval, where the
reaction with constant $k$ (bold arrow) is situated. The right
eigenvector $r$ has coordinate 1 for the vertex, which is the
beginning of the reaction with constant $k$, and coordinate $-1$
for the vertex, which is end of the interval in the direction of
reactions. The left eigenvector $l$ has coordinate 1 for the
beginning of the reaction with constant $k$ and for all preceding
vertices from the connected interval. All other coordinates of $r$
and $l$ are zero.}
\end{figure}

If the rate constants any two constants, $k_i$, $k_j$ are connected
by relation $k_i \gg k_j$ or $k_i \ll k_j$ (i.e. they are well
separated), then
\begin{equation}\label{firstMultiL}
 \frac{k_{i-j}}{k_{i-j}-k_i}
 \approx \left\{
    \begin{aligned}
        & 1, \; &\mbox{if} \; k_i \ll k_{i-j};
        \\
        & 0, \;&  \mbox{if} \; k_i \gg k_{i-j},
    \end{aligned}\right.
\end{equation}
Hence, $|l^i_{i-m}| \approx 1$ or $|l^i_{i-m}| \approx 0$. To
demonstrate that also $|r^i_{i+m}| \approx 1$ or $|r^i_{i+m}|
\approx 0$, we shift nominators in the product (\ref{ChainEigen}) on
such a way: $$ r^i_{i+m}= \frac{k_i}{k_{i+m}-k_i} \prod_{j=1}^{m-1}
\frac{k_{i+j}}{k_{i+j}-k_i}.$$ Exactly as in (\ref{firstMultiL}),
each multiplier $\frac{k_{i+j}}{k_{i+j}-k_i}$ here is either almost
1 or almost 0, and $\frac{k_i}{k_{i+m}-k_i}$ is either almost 0 or
almost $-1$. In this zero-one asymptotics
\begin{equation}\label{01cyrcleAsy}
\begin{split}
l^i_i= &1, \; l^i_{i-m}\approx 1 \; \\ &\mbox{if} \; k_{i-j}>k_i
\; \mbox{for all} \; j=1, \ldots m, \; \mbox{else} \;
l^i_{i-m}\approx 0;
\\
r^i_i=&1, \; r^i_{i+m}\approx -1  \; \\& \mbox{if} \; k_{i+j} >k_i
\; \mbox{for all} \; j=1, \ldots m-1 \; \\ &\mbox{and} \; k_{i+m}
< k_i, \;  \mbox{else} \; r^i_{i+m}\approx 0 .
\end{split}
\end{equation}
In this asymptotic (Fig.~\ref{FigChainEigen}), only two
coordinates of right eigenvector $r^i$ can have nonzero values,
$r^i_i = 1$ and $r^i_{i+m}\approx -1$ where $m$ is the first such
positive integer that $i+m < n$ and $k_{i+m} < k_i$. Such $m$
always exists because $k_n=0$. For left eigenvector $l^i$, $l^i_i
\approx \ldots l^i_{i-q}\approx 1$ and $l^i_{i-q-j}\approx 0$
where $j>0$ and $q$ is the first such positive integer that
$i-q-1>0$ and $k_{i-q-1}<k_i$. It is possible that such $q$ does
not exist. In that case, all $l^i_{i-j}\approx 1$ for $j\geq 0$.
It is straightforward to check that in this asymptotic $l^i r^j =
\delta_{ij}$.

The simplest example gives the order $k_1 \gg k_{2} \gg ... \gg
k_{n-1}$: $l^i_{i-j}\approx 1$ for $j\geq 0$, $r^i_i = 1$,
$r^i_{i+1}\approx -1$ and all other coordinates of eigenvectors
are close to zero. For the inverse order, $k_1 \ll k_{2} \ll ...
\ll k_{n-1}$, $l^i_{i} = 1$, $r^i_i = 1$, $r^i_{n}\approx -1$  and
all other coordinates of eigenvectors are close to zero.

For less trivial example, let us find the asymptotic of left and
right eigenvectors for a chain of reactions:
 $$A_1 {\rightarrow^{\!\!\!\!\!\!5}}\,\,
A_2{\rightarrow^{\!\!\!\!\!\!3}}\,\,
A_3{\rightarrow^{\!\!\!\!\!\!4}}\,\,
A_4{\rightarrow^{\!\!\!\!\!\!1}}\,\,
A_5{\rightarrow^{\!\!\!\!\!\!2}}\,\, A_6 ,$$
 where  the upper index marks the order of rate constants:
 $k_4 \gg k_5 \gg k_2\gg k_3\gg k_1$ ($k_i$ is the
 rate constant of reaction $A_i \to ...$).

 For left eigenvectors, rows $l^i$, we have the
following asymptotics:
\begin{equation}\label{LeftApproxLin}
\begin{split}
&l^1\approx (1,0,0,0,0,0), \;  l^2\approx (0,1,0,0,0,0), \; \\
&l^3\approx (0,1,1,0,0,0),  l^4\approx (0,0,0,1,0,0), \; \\
&l^5\approx (0,0,0,1,1,0).
\end{split}
\end{equation}
For right eigenvectors, columns $r^i$, we have the following
asymptotics (we write  vector-columns in rows):
\begin{equation}\label{RightApproxLin}
\begin{split}
&r^1\approx
 (1, 0, 0, 0, 0, -1), \;
r^2\approx
 (0, 1, -1, 0, 0, 0), \; \\
&r^3\approx
 (0, 0, 1, 0, 0, -1),
r^4\approx
 (0, 0, 0, 1, -1, 0), \; \\
&r^5\approx
 (0, 0, 0, 0, 1, -1).
\end{split}
\end{equation}
The corresponding approximation to the general solution of the
kinetic equations is:
\begin{equation}\label{genSol}
c(t)=(l^0, c(0)) r^0 + \sum_{i=1}^{n-1}  (l^i c(0)) r^i \exp(-k_i
t),
\end{equation}
where $c(0)$ is the initial concentration vector, and for left and
right eigenvectors $l^i$ and $r^i$ we use their zero-one
asymptotic. In other words, approximation of the left eigenvectors
provides us with almost exact lumping (for analysis of exact
lumping see the paper by \cite{LumpLiRab1}) .

\section{Acyclic Non-branching Network: Explicit Formulas for Eigenvectors  \label{sec4}}

So, to analyze asymptotic of eigenvalues and eigenvectors for a
irreversible cycle, we cut the reaction with the smallest
constant, get a linear chain, and analyze the eigenvalues and
eigenvectors for this chain. For a general multiscale reaction
network (instead of a cycle) we will come, after some surgery, to
acyclic non-branching reaction networks  (instead of a linear
chain).

For any network without branching, we can simplify the notation
for the kinetic constants, by introducing $\kappa_i = k_{ji}$ for
the only reaction $A_i \to A_j$, or $\kappa_i = 0$, if there is no
such a reaction. Also it is useful to introduce a map $\phi$ on
the set of vertices: $\phi(i)=j$, if there exist reaction $A_i \to
A_j$, and $\phi(i)=i$ if there are no outgoing reactions from the
$A_i \to A_j$. For iterations of the map $\phi$ we use notation
$\phi^q$.

For an acyclic non-branching reaction network, for any vertex
$A_i$ there is an eigenvalue $- \kappa_i$ and the corresponding
eigenvector. If $A_i$ is a sink vertex, then this eigenvalue is
zero. For left and right eigenvectors of $K$ that correspond to
$A_i$ we use notations $l^i$ (vector-row) and $r^i$
(vector-column), correspondingly.

Let us suppose that $A_f$ is a sink vertex of the network. Its
associated right and left eigenvectors corresponding to the zero
eigenvalue are given by: $r^i_j = \delta_{ij}$; $l^i_j=1$ if and
only if $\phi^q(j)=i$ for some $q>0$.

\begin{figure}[t]
\centering{ \includegraphics[width=70mm]{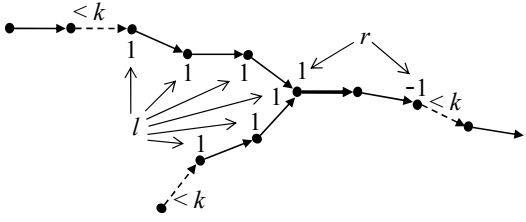} }
\caption{\label{FigTreeEigen} Graphical representation of
eigenvectors approximation for the acyclic non-branching reaction
network with well separated constants (compare to
Fig.~\ref{FigChainEigen}). The eigenvalue $-k$ corresponds to the
reaction $A_i \to A_{\phi(i)}$ (bold arrow). To the right from $A_i$
are vertices $A_{\phi^q(i)}$ and to the left are those $A_j$, for
which there exists such $q$ that $\phi^q(j)=i$. The reactions with
the rate constants $<k$ (dashed lines) are deleted from the network.
The right and left eigenvectors could have nonzero coordinates only
for vertices from the maximal connected subgraph of the presented
graph, where the $A_i$ is situated. The right eigenvector $r$ has
coordinate 1 for $A_i$ (beginning of the bold arrow), and coordinate
$-1$ for the vertex, which is the minimal in that connected
subgraph. The left eigenvector $l$ has coordinate 1 for the
beginning of the reaction with constant $k$ and for all preceding
vertices from the subgraph. All other coordinates of $r$ and $l$ are
zero.}
\end{figure}

For nonzero eigenvalues, right eigenvectors will be constructed by
recurrence starting from the vertex $A_i$ and moving in the
direction of the flow. The construction is in opposite direction for
left eigenvectors.

For right eigenvector $r^i$ only coordinates $r^i_{\phi^k(i)}$
($k=0,1,\ldots \tau_i$) could have nonzero values, and
\begin{equation}\label{rightAcyc}
\begin{split}
r^i_{\phi^{k+1}(i)}=
\frac{\kappa_{\phi^{k}(i)}}{\kappa_{\phi^{k+1}(i)}-\kappa_i}
r^i_{\phi^{k}(i)}= \prod_{j=0}^k
\frac{\kappa_{\phi^{j}(i)}}{\kappa_{\phi^{j+1}(i)}-\kappa_i} \\
 = \frac{\kappa_i}{\kappa_{\phi^{k+1}(i)}-\kappa_i}
 \prod_{j=0}^{k-1}\frac{\kappa_{\phi^{j+1}(i)}}{\kappa_{\phi^{j+1}(i)}-\kappa_i}.
\end{split}
\end{equation}
For left eigenvector $l^i$ coordinate $l^i_j$  could have nonzero
value only if there exists such $q\geq 0$ that $\phi^q(j)=i$ (this
$q$ is unique because the system is acyclic):
\begin{equation}\label{leftAcyc}
l^i_j=\frac{\kappa_j}{\kappa_j - \kappa_i} l^i_{\phi(j)}=
\prod_{k=0}^{q-1} \frac{\kappa_{\phi^k(j)}}{\kappa_{\phi^k(j)} -
\kappa_i}.
\end{equation}

For well separated constants,  we can write the asymptotic
representation explicitly, analogously to (\ref{01cyrcleAsy})
(Fig.~\ref{FigTreeEigen}). For left eigenvectors, $l^i_i=1$ and
$l^i_j=1$ (for $i\neq j$) if there exists such $q$ that
$\phi^q(j)=i$, and $\kappa_{\phi^d(j)}>\kappa_i$ for all $d= 0,
\ldots q-1$, else $l^i_j=0$. For right eigenvectors, $r^i_i=1$ and
$r^i_{\phi^k(i)}=-1$ if $\kappa_{\phi^k(i)}< \kappa_i$ and for all
positive $m<k$ inequality  $\kappa_{\phi^m(i)}> \kappa_i$ holds,
i.e. $k$ is first such positive integer that $\kappa_{\phi^k(i)}<
\kappa_i$ (for fixed point $A_p$ we use $\kappa_p=0$). Vector
$r^i$ has not more than two nonzero coordinates. It is
straightforward to check that in this asymptotic $l^i r^j =
\delta_{ij}$.

For example, let us find that asymptotic for a branched acyclic
system of reactions:
\begin{equation*}
%\begin{split}
A_1 {\rightarrow^{\!\!\!\!\!\!7}}\,\,
A_2{\rightarrow^{\!\!\!\!\!\!5}}\,\,
A_3{\rightarrow^{\!\!\!\!\!\!6}}\,\,
A_4{\rightarrow^{\!\!\!\!\!\!2}}\,\,
A_5{\rightarrow^{\!\!\!\!\!\!4}}\,\, A_8, \;\;
A_6{\rightarrow^{\!\!\!\!\!\!1}}\,\,
A_7{\rightarrow^{\!\!\!\!\!\!3}}\,\,A_4
%\end{split}
\end{equation*}
 where  the upper index marks the order of rate constants:
 $\kappa_6>\kappa_4>\kappa_7>\kappa_5>\kappa_2>\kappa_3>\kappa_1$ ($\kappa_i$ is the
 rate constant of reaction $A_i \to ...$).

For zero eigenvalue, the left and right eigenvectors are $$l^8=
(1,1,1,1,1,1,1,1,1), \; r^8= (0,0,0,0,0,0,0,1).$$ For left
eigenvectors, rows $l^i$, that correspond to nonzero eigenvalues
we have the following asymptotics:
\begin{equation}\label{leftAcycBran}
\begin{split}
&l^1\approx (1,0,0,0,0,0,0,0), \; l^2\approx (0,1,0,0,0,0,0,0), \;
\\
&l^3\approx (0,1,1,0,0,0,0,0), l^4\approx (0,0,0,1,0,0,0,0), \; \\
&l^5\approx (0,0,0,1,1,1,1,0), \; l^6\approx (0,0,0,0,0,1,0,0). \\
&l^7\approx (0,0,0,0,0,1,1,0)
\end{split}
\end{equation}
For the corresponding right eigenvectors, columns $r^i$, we have
the following asymptotics (we write vector-columns in rows):
\begin{equation}
\begin{split}
&r^1 \!\approx \!
 (1, 0, 0, 0, 0, 0, 0, -1),  \,
r^2 \!\approx  \!
 (0, 1, -1, 0, 0, 0, 0, 0 ), \\
&r^3 \!\approx  \!
 (0, 0, 1,  0, 0, 0, 0, -1), \,
r^4 \!\approx  \!
 (0, 0, 0, 1,  -1,0 ,0 ,0 ), \\
&r^5 \!\approx  \!
 (0, 0, 0, 0, 1 ,0 ,0 , -1), \,
r^6 \!\approx  \!
 (0, 0, 0,  0, 0, 1, -1, 0),  \\
&r^7 \!\approx  \!
 (0, 0, 0, 0,  -1, 0 ,1, 0  ).
\end{split}
\end{equation}

\section{Calculating the Dominant System for a Linear Multiscale Network  \label{sec5}}

\subsection{Problem Statement}

We study asymptotical behavior of the transformation of the kinetic
matrix $K$ to the normal form along the lines $\ln k_{ij}=
\theta_{ij} \xi$ when $\xi \to \infty$. For almost all direction
vectors $(\theta_{ij})$ (outside several hyperplanes) there exists a
minimal reaction network which reaction rate constants are monomials
of $k_{ij}$ ($\prod_{ij} k_{ij}^{f_{ij}}$, where $f_{ij}$ are not
obligatory positive numbers) and eigenvectors and eigenvalues
approximate the eigenvectors and eigenvalues when $\xi \to \infty$
with arbitrary high relative accuracy. We call this minimal system
the {\it dominant system}. Existence of dominant systems is proven
by direct construction (this Sec.) and estimates of accuracy of
approximations (Appendix).

The dominant systems coincide for vectors $(\theta_{ij})$ from some
polyhedral cones. Therefore, we don't need to study a given value of
$(\theta_{ij})$ but rather have to build these cones together with
the correspondent dominant systems. The following formal rule
(``assumption of well separated constants") allows us to simplify
this task: if in construction of dominant systems we need to compare
two monomials, $M_f=\prod_{ij} k_{ij}^{f_{ij}}$ and $M_g=\prod_{ij}
k_{ij}^{g_{ij}}$ then we can always state that either $M_f \gg M_g$
or $M_f \ll M_g$ and consider the logarithmic hyperplane $M_f = M_g$
as a boundary between different cones. At the end, we can join all
cones with the same dominant system. We are interested in robust
asymptotic and do not analyze directions $(\theta_{ij})$ which
belong to the boundary hyperplanes. This robust asymptotic with well
separated constants and acyclic dominant systems is typical because
the exclusive direction vectors belon to a finite number of
hyperplanes.

There may be other approaches based on (i) the Maslov dequantization
and idempotent algebras (\cite{LitMas}), (ii) the limit of
log-uniform distributions in wide boxes of constants under some
conditions (\cite{Rabi3,GorbaRadul2008}), or (iii) on consideration
of all possible orderings of all monomials with integer exponents
and construction of correspondent dominant systems (\cite{orderRobb}
proved that there exists only a final number of such orderings and
enumerated all of them, see also the book by \cite{orderPfi}). They
give the same final result but with different intermediate steps.

\subsection{Auxiliary Operations}
\subsubsection{From Reaction Network to Auxiliary Dynamical
System}

Let us consider a reaction network $\mathcal{W}$ with a given
structure and fixed ordering of constants. The set of vertices of
$\mathcal{W}$ is $\mathcal{A}$ and the set of elementary reactions
is $\mathcal{R}$. Each reaction from $\mathcal{R}$ has the form
$A_i \to A_j$, $A_i, A_j \in \mathcal{A}$. The corresponding
constant is $k_{ji}$. For each $A_i \in \mathcal{A}$ we define
$\kappa_i= \max_j \{k_{ji} \}$ and $\phi (i) = {\rm arg \, max}_j
\{k_{ji} \}$. In addition, $\phi (i)=i$ if $k_{ji}=0$ for all $j$.

\begin{figure}
\centering{ \includegraphics[width=30mm]{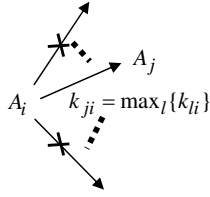} }
\caption{\label{auxcut} Construction of the auxiliary reaction
network by pruning. For every vertex, it is necessary to leave the
outgoing reaction with maximal reaction rate constant. Other
reactions should be deleted.}
\end{figure}

The {\it auxiliary discrete dynamical system} for the reaction
network $\mathcal{W}$ is the dynamical system
$\Phi=\Phi_{\mathcal{W}}$ defined by the map $\phi$ on the finite
set $\mathcal{A}$. The {\it auxiliary reaction network}
(Fig.~\ref{auxcut}) $\mathcal{V}=\mathcal{V}_{\mathcal{W}}$ has
the same set of vertices $\mathcal{A}$ and the set of reactions
$A_i \to A_{\phi(i)}$ with reaction constants $\kappa_i$.
Auxiliary kinetics is described by $\dot c = \tilde K c$, where
$\tilde {K}_{ij}= - \kappa_j \delta_{ij} +  \kappa_j \delta_{i \,
\phi (j)} $.

\subsubsection{Decomposition of Discrete Dynamical Systems on
Finite Sets}

Discrete dynamical system on a finite set $V= \{A_1,A_2, \ldots
A_n \}$ is a semigroup $1, \phi, \phi^2, ...$, where $\phi$ is a
map $\phi: V \to V$.  $A_i \in V$ is a periodic point, if $\phi^l
(A_i)=A_i$ for some $l>0$; else $A_i$ is a transient point. A
cycle of period $l$ is a sequence of $l$ distinct periodic points
$A, \phi(A), \phi^2(A), \ldots \phi^{l-1}(A)$ with $\phi^{l}(A) =
A$. A cycle of period one consists of one fixed point,
$\phi(A)=A$. Two cycles, $C, C'$ either coincide or have empty
intersection.

The set of periodic points, $V^{\rm p}$, is always nonempty. It is
a union of cycles: $V^{\rm p}= \cup_j C_j$. For each point $A \in
V$ there exist such a positive integer $\tau(A)$ and a cycle
$C(A)=C_j$ that $\phi^q(A) \in C_j$ for $q \geq \tau(A)$. In that
case we say that $A$ belongs to basin of attraction of cycle $C_j$
and use notation $Att(C_j)= \{A\ | \ C(A) = C_j \}$. Of course,
$C_j \subset Att(C_j)$. For different cycles, $Att(C_j) \cap
Att(C_l) = \varnothing$. If $A$ is periodic point then
$\tau(A)=0$. For transient points $\tau(A)>0$.

\begin{figure}
\centering{ \includegraphics[width=60mm]{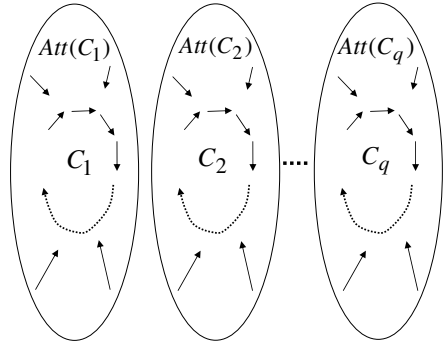}}
\caption{\label{DecompDiscrDynSys}Decomposition of a discrete
dynamical system.}
\end{figure}

So, the phase space $V$ is divided onto subsets $Att(C_j)$
(Fig.~\ref{DecompDiscrDynSys}). Each of these subsets includes one
cycle (or a fixed point, that is a cycle of length 1). Sets
$Att(C_j)$ are $\phi$-invariant: $\phi(Att(C_j)) \subset Att(C_j)
$. The set $Att(C_j) \setminus C_j$ consist of transient points
and there exists such positive integer $\tau$ that
$\phi^q(Att(C_j))=C_j$ if $q \geq \tau$.

Discrete dynamical systems on a finite sets correspond to graphs
without branching points. Notice that for the graph that
represents a discrete dynamic system, attractors are ergodic
components, while basins are connected components.

\subsection{Algorithm for Calculating the Dominant
System}

For this general case, the algorithm consists of two main
procedures: (i) cycles gluing and (ii) cycles restoration and
cutting.

\subsubsection{Cycles Gluing}
Let us start from a reaction network $\mathcal{W}$ with a given
structure and fixed ordering of constants. The set of vertices of
$\mathcal{W}$ is $\mathcal{A}$ and the set of elementary reactions
is $\mathcal{R}$.

If all attractors of the auxiliary dynamic system
$\Phi_{\mathcal{W}}$ are fixed points $A_{f1}, A_{f2}, ... \in
\mathcal{A}$, then  the auxiliary reaction network is acyclic, and
the auxiliary kinetics approximates relaxation of the whole
network $\mathcal{W}$.

In general case, let the system $\Phi_{\mathcal{W}}$ have several
attractors that are not fixed points, but cycles $C_1,C_2, ...$
with periods $\tau_1, \tau_2,...>1$. By gluing these cycles in
points, we transform the reaction network $\mathcal{W}$ into
$\mathcal{W}^1$. The dynamical system $\Phi_{\mathcal{W}}$ is
transformed into $\Phi^1$.  For these new system and network, the
connection $\Phi^1=\Phi_{\mathcal{W}^1}$ persists: $\Phi^1$ is the
auxiliary discrete dynamical system for $\mathcal{W}^1$.

For each cycle, $C_i$, we introduce a new vertex $A^i$. The new
set of vertices, $\mathcal{A}^1=\mathcal{A} \cup\{A^1,A^2,...\}
\setminus (\cup_i C_i) $ (we delete cycles $C_i$ and add vertices
$A^i$).

\begin{figure}
\centering{ \includegraphics[width=55mm]{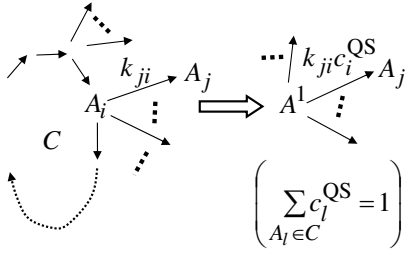}}
\caption{\label{GluCircle}Gluing a cycle with rate constants
renormalization. $c^{\rm QS}_l$ are the quasistationary
concentrations on the cycle. After gluing, we have to leave the
outgoing from $A^1$ reaction with the maximal renormalized rate
constant, and delete others.}
\end{figure}

All the reaction $A \to B$ from the initial set $\mathcal{R}$,
($A, B \in \mathcal{A}$) can be separated into 5 groups:
\begin{enumerate}
\item{both $A,B \notin \cup_i C_i$;}
\item{$A \notin \cup_i C_i$, but $B \in C_i$;}
\item{$A\in C_i$, but $B \notin \cup_i C_i$;}
\item{$A \in C_i$, $B \in C_j$, $i \neq j$;}
\item{$A,B \in C_i$.}
\end{enumerate}
Reactions from the first group do not change. Reaction from the
second group transforms into $A\to A^i$ (to the whole glued cycle)
with the same constant. Reaction of the third type changes into
$A^i \to B$ with the {\it rate constant renormalization}: let the
cycle $C^i$ be the following sequence of reactions $A_{1} \to
A_{2} \to ... A_{\tau_i} \to A_1$, and the reaction rate constant
for $A_i \to A_{i+1}$ is $k_i$ ($k_{\tau_i}$ for $A_{\tau_i} \to
A_1$). For the limiting reaction of the cycle $C_i$ we use
notation $k_{\lim \, i}$. If $A=A_j$ and $k$ is the rate reaction
for $A \to B$, then the new reaction $A^i \to B$ has the rate
constant $k k_{\lim \, i}/ k_j$. This corresponds to a
quasistationary distribution on the cycle (\ref{CycleLimRateLin}).
The new rate constant is smaller than the initial one: $k k_{\lim
\, i}/ k_j < k$, because  $k_{\lim \, i} < k_j $ due to definition
of limiting constant. The same constant renormalization is
necessary for reactions of the fourth type. These reactions
transform into $A^i \to A^j$. Finally, reactions of the fifth type
vanish.

After we glue all the cycles (Fig.~\ref{GluCircle}) of auxiliary
dynamical system in the reaction network $\mathcal{W}$, we get
$\mathcal{W}^1$. Let us assign $\mathcal{W}:=\mathcal{W}^1$,
$\mathcal{A}:=\mathcal{A}^1$ and iterate until we obtain an acyclic
network and exit. This acyclic network is a ``forest" and consists
of trees oriented from leafs to a root. The number of such trees
coincide with the number of fixed points in the final network.

After gluing we can identify the reactions, which will be included
into the dominant system. Their constants are the {\it critical
parameters of the networks}. The list of these parameters,
consists of all reaction rates of the final acyclic auxiliary
network, and of the rate constants of the glued cycles, but
without their limiting steps. Some of these parameters are rate
constants of the initial network, other have the monomial
structure. Other constants and corresponding reactions do not
participate in the following operations. To form the structure of
the dominant network, we need one more procedure.

\subsubsection{Cycles Restoration and Cutting \label{SubsecRestore}}

We start the reverse process from the glued network $\mathcal{V}^m$
on $\mathcal{A}^m$. On a step back, from the set $\mathcal{A}^m$ to
$\mathcal{A}^{m-1}$ and so on, some of glued cycles should be
restored and cut. On the $q$th step we build an acyclic reaction
network on $\mathcal{A}^{m-q}$, the final network is defined on the
initial vertex set and approximates relaxation of $\mathcal{W}$.

To make one step back from $\mathcal{V}^m$ let us select the
vertices of $\mathcal{A}^m$ that are glued cycles from
$\mathcal{V}^{m-1}$. Let these vertices be $A^m_{1}, A^m_{2},
...$. Each $A^m_i$ corresponds to a glued cycle from
$\mathcal{V}^{m-1}$, $A^{m-1}_{i1} \to A^{m-1}_{i2} \to ...
A^{m-1}_{i \tau_i} \to A^{m-1}_{i1}$, of the length $\tau_i$. We
assume that the limiting steps in these cycles are $A^{m-1}_{i
\tau_i} \to A^{m-1}_{i1}$. Let us substitute each vertex $A^m_i$
in $\mathcal{V}^m$ by $\tau_i$ vertices $A^{m-1}_{i1} ,
A^{m-1}_{i2}, ... A^{m-1}_{i\tau_i}$ and add to $\mathcal{V}^m$
reactions $A^{m-1}_{i1} \to A^{m-1}_{i2} \to ... A^{m-1}_{i
\tau_i}$ (that are the cycle reactions without the limiting step)
with corresponding constants from $\mathcal{V}^{m-1}$.

\begin{figure}[t]
\centering{ \includegraphics[width=70mm]{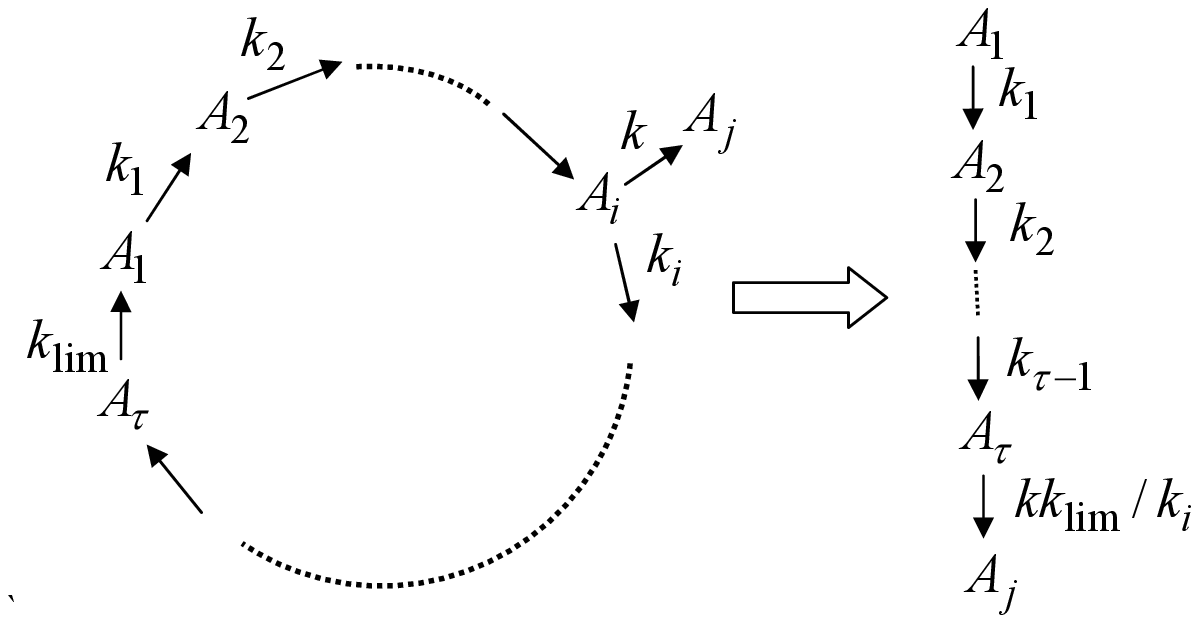} }
\caption{\label{CycleSurg} The main operation of the cycle surgery:
on a step back we get a cycle $A_{1} \to... \to A_{\tau} \to A_{1}$
with the limiting step $A_{\tau} \to A_{1}$ and one outgoing
reaction $A_{i} \to A_j$. We should delete the limiting step,
reattach (``recharge") the outgoing reaction $A_{i} \to A_j$ from
$A_i$ to $A_{\tau}$ and change its rate constant $k$ to the rate
constant $k k_{\lim}/k_{i}$. The new value of reaction rate constant
is always smaller than the initial one: $k k_{\lim}/k_{i} < k$ if
$k_{\lim}\neq k_{i}$. For this operation only one condition $k \ll
k_i$ is necessary ($k$ should be small with respect to reaction
$A_{i}\to A_{i+1}$ rate constant, and can exceed any other reaction
rate constant).}
\end{figure}

If there exists an outgoing reaction $A^m_i \to B$ in
$\mathcal{V}^m$ then we substitute it by the reaction $A^{m-1}_{i
\tau_i} \to B$ with the same constant, i.e. outgoing reactions
$A^m_i \to...$ are reattached to the heads of the limiting steps
(Fig.~\ref{CycleSurg}). Let us rearrange reactions from
$\mathcal{V}^m$ of the form $B \to A^m_i$. These reactions have
prototypes in $\mathcal{V}^{m-1}$ (before the last gluing). We
simply restore these reactions. If there exists a reaction $A^m_i
\to A^m_j$ then we find the prototype in $\mathcal{V}^{m-1}$, $A \to
B$, and substitute the reaction by $A^{m-1}_{i \tau_i} \to B$ with
the same constant, as for $A^m_i \to A^m_j$.

After that step is performed, the vertices set is
$\mathcal{A}^{m-1}$, but the reaction set differs from the reactions
of the network $\mathcal{V}^{m-1}$: the limiting steps of cycles are
excluded and the outgoing reactions of glued cycles are included
(reattached to the heads of the limiting steps). To make the next
step, we select vertices of $\mathcal{A}^{m-1}$ that are glued
cycles from $\mathcal{V}^{m-2}$, substitute these vertices by
vertices of cycles, delete the limiting steps, attach outgoing
reactions to the heads of the limiting steps, and for incoming
reactions restore their prototypes from $\mathcal{V}^{m-2}$, and so
on.

After all, we restore all the glued cycles, and construct an acyclic
reaction network on the set $\mathcal{A}$. This acyclic network
approximates relaxation of the network $\mathcal{W}$. We call this
system the dominant system of $\mathcal{W}$ and use notation ${\rm
dom \, mod}(\mathcal{W})$.

In the simplest case, the dominant system is determined by the
ordering of constants. But for sufficiently complex systems we
need to introduce auxiliary elementary reactions. They appear
after cycle gluing and have monomial rate constants of the form
$k_{\varsigma}=\prod_i k_i^{\varsigma_i}$, where ${\varsigma_i}$
are integers, but not mandatory positive. The dominant system
depends on the place of these monomial values among the ordered
constants. For systems with well separated constants we can also
assume that each of these new constants will be well separated
from other constants (\cite{GorbaRadul2008}).

\subsection{Example}

To demonstrate a possible branching of described algorithm for
cycles surgery (gluing, restoring and cutting) with necessity of
additional orderings, let us consider the following system:

\begin{equation}\label{chain+-}
A_1 {\rightarrow^{\!\!\!\!\!\!1}}\,\,
A_2{\rightarrow^{\!\!\!\!\!\!6}}\,\,
A_3{\rightarrow^{\!\!\!\!\!\!2}}\,\,
A_4{\rightarrow^{\!\!\!\!\!\!3}}\,\,
A_5{\rightarrow^{\!\!\!\!\!\!4}}\,\, A_3, \;\;
A_4{\rightarrow^{\!\!\!\!\!\!5}}\,\, A_2, \;\;
\end{equation}
(where  the upper index marks the order of rate constants). The
auxiliary discrete dynamical system for reaction network
(\ref{chain+-}) is $$A_1 {\rightarrow^{\!\!\!\!\!\!1}}\,\,
A_2{\rightarrow^{\!\!\!\!\!\!6}}\,\,
A_3{\rightarrow^{\!\!\!\!\!\!2}}\,\,
A_4{\rightarrow^{\!\!\!\!\!\!3}}\,\,
A_5{\rightarrow^{\!\!\!\!\!\!4}}\,\, A_3.$$
 It has only one attractor, a cycle
$A_3{\rightarrow^{\!\!\!\!\!\!2}}\,\,
A_4{\rightarrow^{\!\!\!\!\!\!3}}\,\,
A_5{\rightarrow^{\!\!\!\!\!\!4}}\,\, A_3$. This cycle is not a sink
for the whole network (\ref{chain+-}) because reaction
$A_4{\rightarrow^{\!\!\!\!\!\!5}}\,\, A_2$ leads from that cycle.
After gluing the cycle into a vertex $A^1_3$ we get the new network
$A_1 {\rightarrow^{\!\!\!\!\!\!1}}\,\,
A_2{\rightarrow^{\!\!\!\!\!\!6}}\,\,
A^1_3{\rightarrow^{\!\!\!\!\!\!?}}\,\,A_2$. The rate constant for
the reaction $A^1_3{\rightarrow}A_2$ is $k^1_{23}=
k_{24}k_{35}/k_{54}$, where $k_{ij}$ is the rate constant for the
reaction $A_j \to A_i$ in the initial network ($k_{35}$ is the cycle
limiting reaction). The new network coincides with its auxiliary
system and has one cycle, $A_2{\rightarrow^{\!\!\!\!\!\!6}}\,\,
A^1_3{\rightarrow^{\!\!\!\!\!\!?}}\,\,A_2$. This cycle is a sink,
hence, we can start the back process of cycles restoring and
cutting. One question arises immediately: which constant is smaller,
$k_{32}$ or $k^1_{23}$. The smallest of them is the limiting
constant, and the answer depends on this choice. Let us consider two
possibilities separately: (1) $k_{32} > k^1_{23}$ and (2) $k_{32} <
k^1_{23}$.

(1) Let as assume that $k_{32}
> k^1_{23}$. The final auxiliary system after gluing cycles is $A_1
{\rightarrow^{\!\!\!\!\!\!1}}\,\,
A_2{\rightarrow^{\!\!\!\!\!\!6}}\,\,
A^1_3{\rightarrow^{\!\!\!\!\!\!?}}\,\,A_2$. Let us delete  the
limiting reaction $A^1_3{\rightarrow^{\!\!\!\!\!\!?}}\,\,A_2$ from
the cycle. We get an acyclic system $A_1
{\rightarrow^{\!\!\!\!\!\!1}}\,\,
A_2{\rightarrow^{\!\!\!\!\!\!6}}\,\, A^1_3$. The component $A^1_3$
is the glued cycle $A_3{\rightarrow^{\!\!\!\!\!\!2}}\,\,
A_4{\rightarrow^{\!\!\!\!\!\!3}}\,\,
A_5{\rightarrow^{\!\!\!\!\!\!4}}\,\, A_3$. Let us restore this cycle
and delete the limiting reaction
$A_5{\rightarrow^{\!\!\!\!\!\!4}}\,\, A_3$. We get the dominant
system $A_1 {\rightarrow^{\!\!\!\!\!\!1}}\,\,
A_2{\rightarrow^{\!\!\!\!\!\!6}}\,\,A_3{\rightarrow^{\!\!\!\!\!\!2}}\,\,
A_4{\rightarrow^{\!\!\!\!\!\!3}}\,\, A_5$. Relaxation of this system
approximates relaxation of the initial network (\ref{chain+-}) under
additional condition $k_{32} > k^1_{23}$.

(2) Let as assume now that  $k_{32} < k^1_{23}$. The final auxiliary
system after gluing cycles is the same, $A_1
{\rightarrow^{\!\!\!\!\!\!1}}\,\,
A_2{\rightarrow^{\!\!\!\!\!\!6}}\,\,
A^1_3{\rightarrow^{\!\!\!\!\!\!?}}\,\,A_2$, but the limiting step in
the cycle is different, $A_2{\rightarrow^{\!\!\!\!\!\!6}}\,\,
A^1_3$. After cutting this step, we get acyclic system $A_1
{\rightarrow^{\!\!\!\!\!\!1}}\,\, A_2{\leftarrow^{\!\!\!\!?}}\,
A^1_3$, where the last reaction has rate constant $k^1_{23}$.

The component $A^1_3$ is the glued cycle
$$A_3{\rightarrow^{\!\!\!\!\!\!2}}\,\,
A_4{\rightarrow^{\!\!\!\!\!\!3}}\,\,
A_5{\rightarrow^{\!\!\!\!\!\!4}}\,\, A_3\, .$$ Let us restore this
cycle and delete the limiting reaction
$A_5{\rightarrow^{\!\!\!\!\!\!4}}\,\, A_3$. The connection from
glued cycle $A^1_3{\rightarrow^{\!\!\!\!\!\!?}}\,\,A_2$ with
constant $k^1_{23}$ transforms into connection
$A_5{\rightarrow^{\!\!\!\!\!\!?}}\,\,A_2$ with the same constant
$k^1_{23}$.

We get the dominant system: $$A_1 {\rightarrow^{\!\!\!\!\!\!1}}\,\,
A_2\, , \; A_3{\rightarrow^{\!\!\!\!\!\!2}}\,\,
A_4{\rightarrow^{\!\!\!\!\!\!3}}\,\, A_5
{\rightarrow^{\!\!\!\!\!\!?}}\,\,A_2\, .$$ The order of constants is
now known: $k_{21}>k_{43}>k_{54}>k^1_{23}$, and we can substitute
the sign ``?" by ``4": $A_3{\rightarrow^{\!\!\!\!\!\!2}}\,\,
A_4{\rightarrow^{\!\!\!\!\!\!3}}\,\, A_5
{\rightarrow^{\!\!\!\!\!\!4}}\,\,A_2$.

For both cases, $k_{32} > k^1_{23}$ ($k^1_{23}=
k_{24}k_{35}/k_{54}$) and $k_{32} < k^1_{23}$ it is easy to find the
eigenvectors explicitly  and to write the solution to the kinetic
equations in explicit form.

\section{The Reversible Triangle of Reactions \label{sec6}}

In this section, we illustrate the analysis of dominant systems on
a simple example, the reversible triangle of reactions.
\begin{equation}\label{revtri}
A_1 \leftrightarrow A_2 \leftrightarrow A_3 \leftrightarrow A_1 \,
\end{equation}
This triangle appeared in many works as an ideal object for a case
study. Our favorite example is the work of \cite{Wei62}. Now in
our study the triangle (\ref{revtri}) is  not necessarily a closed
system. We can assume that it is a subsystem of a larger system,
and any reaction $A_i \to A_j$ represents a reaction of the form
$\ldots +A_i \to A_j + \ldots$, where unknown but slow components
are substituted by dots. This means that there are no mandatory
relations between reaction rate constants, and six reaction rate
constants are arbitrary nonnegative numbers.

Let the reaction rate constant $k_{21}$ for the reaction $A_1 \to
A_2$ be the largest.

\begin{figure}
\centering{
\includegraphics[width=75mm]{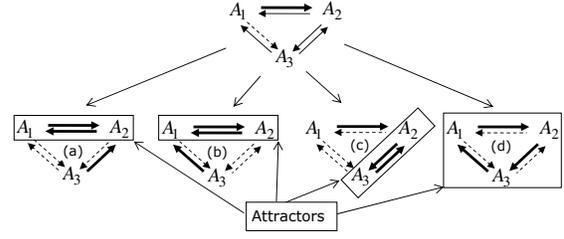}
\caption{\label{TriangleAux} Four possible auxiliary dynamical
systems for the reversible triangle of reactions with $k_{21} >
k_{ij}$ for $(i,j) \neq (2,1)$: (a) $k_{12} > k_{32}$, $k_{23} >
k_{13}$; (b) $k_{12}
> k_{32}$, $k_{13} > k_{23}$; (c) $k_{32} > k_{12}$, $k_{23} >
k_{13}$; (d) $k_{32} > k_{12}$, $k_{13} > k_{23}$. For each vertex
the outgoing reaction with the largest rate constant is
represented by the solid bold arrow, and other reactions are
represented by the dashed arrows. The digraphs formed by solid
bold arrows are the auxiliary discrete dynamical systems.
Attractors of these systems are isolated in frames. }}
\end{figure}

Let us describe all possible auxiliary dynamical systems for the
triangle (\ref{revtri}). For each vertex, we have to select the
fastest outgoing reaction. For $A_1$, it is always $A_1 \to A_2$,
because of our choice of enumeration (the higher scheme in
Fig.~\ref{TriangleAux}). There exist two choices of the fastest
outgoing reaction for two other vertices and, therefore, only four
versions of auxiliary dynamical systems for (\ref{revtri})
(Fig.~\ref{TriangleAux}). Let us analyze in detail case (a). For
the cases (b) and (c) the details of computations are similar. The
irreversible cycle (d) is even simpler and was already discussed.

\subsection{Auxiliary System (a): $A_1 \leftrightarrow A_2
\leftarrow A_3$; $k_{12} > k_{32}$, $k_{23} > k_{13}$}

\subsubsection{Gluing Cycles}

The attractor is a cycle (with only two vertices) $A_1
\leftrightarrow A_2$. This is not a sink, because two outgoing
reactions exist: $A_1 \to A_3$ and $A_2 \to A_3$. They are
relatively slow: $k_{31} \ll k_{21}$ and $k_{32} \ll k_{12}$. The
limiting step in this cycle is $A_2 \to A_1$ with the rate
constant $k_{12}$. We have to glue the cycle $A_1 \leftrightarrow
A_2$ into one new component $A_1^1$ and to add a new reaction
$A_1^1 \rightarrow A_3$ with the rate constant (see
Fig.~\ref{GluCircle})
\begin{equation}\label{gluedcona}
k_{31}^1=\max\{k_{32}, \, k_{31}k_{12}/k_{21}\}\, .
\end{equation}

As a result, we get a new system, $A_1^1 \leftrightarrow A_3$ with
reaction rate constants $k_{31}^1$ (for $A_1^1 \rightarrow A_3$)
and initial $k_{23}$ (for $A_1^1 \leftarrow A_3$). This cycle is a
sink, because it has no outgoing reactions (the whole system is a
trivial example of a sink).

\subsubsection{Dominant System}

At the next step, we have to restore and cut the cycles. First
cycle to cut is the result of cycle gluing, $A_1^1 \leftrightarrow
A_3$. It is necessary to delete the limiting step, i.e. the
reaction with the smallest rate constant. If $k_{31}^1 > k_{23}$,
then we get $A_1^1 \rightarrow A_3$. If, inverse, $k_{23} >
k_{31}^1$, then we obtain $A_1^1 \leftarrow A_3$.

\begin{figure}
\centering{
\includegraphics[width=75mm]{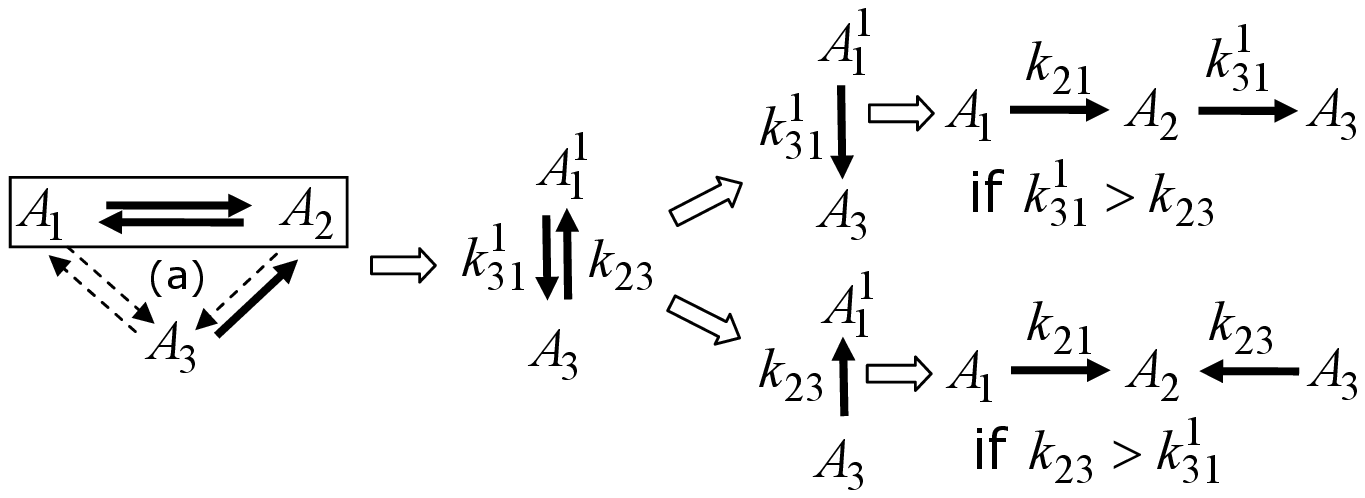}
\caption{\label{Dom1} Dominant systems for case (a) (defined in
Fig.~\ref{TriangleAux})}}
\end{figure}

After that, we have to restore and cut the cycle which was glued
into the vertex $A_1^1$. This is the two-vertices cycle $A_1
\leftrightarrow A_2$. The limiting step for this cycle is $A_1
\leftarrow A_2$, because $k_{21} \gg k_{12}$. If $k_{31}^1 >
k_{23}$, then following the rule visualized by
Fig.~\ref{CycleSurg}, we get the dominant system $A_1 \to A_2 \to
A_3$ with reaction rate constants  $k_{21}$ for $A_1 \to A_2$ and
$k_{31}^1$ for  $A_2 \to A_3$. If $k_{23} > k_{31}^1$ then we
obtain $A_1 \to A_2 \leftarrow A_3$ with reaction rate constants
$k_{21}$ for $A_1 \to A_2$ and $k_{23}$ for $A_2 \leftarrow A_3$.
All the procedure is illustrated by Fig.~\ref{Dom1}.

\subsubsection{Eigenvalues and Eigenvectors}

The eigenvalues and the corresponding eigenvectors for dominant
systems in case (a) are represented below in zero-one asymptotic.
\begin{enumerate}
\item{$k_{31}^1 > k_{23}$,

the dominant system $A_1 \to A_2 \to A_3$,
\begin{equation}
\begin{array}{lll}
  \lambda_0 = 0\, , \;  &r^0\approx(0,0,1)\, , \;
 & l^0=(1,1,1)\, ; \\
  \lambda_1 \approx -k_{21}\, , \; & r^1\approx(1,-1,0)\, , \;
 &l^1\approx(1,0,0)\, ; \\
  \lambda_2 \approx -k_{31}^1\, , \;  & r^2\approx(0,1,-1)\, , \;
 &l^2\approx(1,1,0)\,;
\end{array}
\end{equation}}
\item{$k_{23} > k_{31}^1 $,

the dominant system $A_1 \to A_2 \leftarrow A_3$,
\begin{equation}
\begin{array}{lll}
  \lambda_0 = 0\, , \; & r^0\approx(0,1,0)\, , \;
 &l^0=(1,1,1)\, ; \\
  \lambda_1 \approx -k_{21}\, , \; & r^1\approx(1,-1,0)\, , \;
 &l^1\approx(1,0,0)\, ; \\
  \lambda_2 \approx -k_{23}\, , \; &  r^2\approx(0,-1,1)\, , \;
 &l^2\approx(0,0,1)\,.
\end{array}
\end{equation}}
\end{enumerate}
Here, the value of $k_{31}^1$ is given by formula
(\ref{gluedcona}).

Analysis of examples provided us by an important conclusion: the
number of different dominant systems in examples was less than the
number of all possible orderings. For many pairs of constants
$k_{ij}, k_{lr}$ it is not important which of them is larger. There
is no need to consider all orderings of monomials. We have to
consider only those inequalities between constants and monomials
that appear in the construction of the dominant systems.

\section{Corrections to Dominant Dynamics  \label{sec7}}

The hierarchy of systems $\mathcal{W}$, $\mathcal{W}^1$,
$\mathcal{W}^2$, ... can be used for multigrid correction of the
dominant dynamics. The simple example of multigrid approach gives
the algorithm of steady state approximation (\cite{GorbaRadul2008}).
For this purpose, on the way up (cycle restoration and cutting,
Sec.~\ref{SubsecRestore}) we calculate distribution in restoring
cycles with higher accuracy, by  exact formula  (\ref{CycleRate}),
or in linear approximation (\ref{CycleLimRateLin}) instead of the
simplest zero-one asymptotic (\ref{CycleLimZero}). Essentially, the
way up remains the same.

After termination of the gluing process, we can find all steady
state distributions by restoring cycles in the auxiliary reaction
network $\mathcal{V}^m$. Let $A^m_{f1}, A^m_{f2}, ...$ be fixed
points of $\Phi^m$. The set of steady states for $\mathcal{V}^m$ is
the set of all distributions on the set of fixed points $\{A^m_{f1},
A^m_{f2}, ...\}$.

Let us take one of the basis distributions, $c^m_{fi}=1$, other
$c_i=0$ on $\mathcal{V}^m$. If the vertex $A^m_{fi}$ is a glued
cycle, then we substitute them  by all the vertices of this cycle.
Redistribute the concentration $c^m_{fi}$ between the vertices of
the corresponding cycle by the rule (\ref{CycleRate}) (or by an
approximation). As a result, we get a set of vertices and a
distribution on this set of vertices. If among these vertices there
are glued cycles, then we repeat the procedure of cycle restoration.
Terminate when there is no glued cycles in the support of the
distribution.

The resulting distribution is the approximation to a steady state of
$\mathcal{W}$, and the basis of steady states for $\mathcal{W}$ can
be approximated by this method.

For example, for the system Fig.~\ref{Dom1} we have, first of all,
to compute the stationary distribution in the cycle $A_1^1
\leftrightarrow A_3$, $c^1_1$ and $c_3$. On the base of the general
formula for a simple cycle (\ref{CycleRate}) we obtain:
\begin{equation}\label{fstcyca}
w=\frac{1}{\frac{1}{k_{31}^1} + \frac{1}{k_{23}}}\, , \;
c^1_1=\frac{w}{k_{31}^1}\, , \; c_3=\frac{w}{k_{23}}\, .
\end{equation}

After that, we have to restore the cycle glued into $A_1^1$. This
means to calculate the concentrations of $A_1$ and $A_2$ with
normalization $c_1 + c_2 = c^1_1$. Formula (\ref{CycleRate}) gives:
\begin{equation}\label{scncyca}
w'=\frac{c^1_1}{\frac{1}{k_{21}} +\frac{1}{k_{12}}}\, , \;
c_1=\frac{w'}{k_{21}}\, , \; c_2 =\frac{w'}{k_{12}}\, .
\end{equation}

For eigenvectors, there appear two operations of corrections: (i)
correction for an acyclic network without branching
(\ref{Correction}), (\ref{eigenvectorcorrection}), and (ii)
corrections for a cycle with relatively slow outgoing reactions
(\ref{circlecorrection}). These corrections are by-products of the
accuracy estimates given in Appendix.

\section{Conclusion \label{sec8}}

Now, the idea of limiting step is developed to the asymptotology
of multiscale reaction networks. We found the main terms of
eigenvectors and eigenvalues asymptotic on logarithmic straight
lines $\ln k_{ij}= \theta_{ij} \xi$ when $\xi \to \infty$.  These
main terms could be represented by acyclic dominant system which
is a piecewise constant function of the direction vectors
$(\theta_{ij})$. This theory gives the analogue of the
\cite{ViLju} theory for chemical reaction networks. We
demonstrated also how to construct the accuracy estimates and the
first order corrections to eigenvalues and eigenvectors.

There are several ways of using the developed theory and
algorithms:
\begin{itemize}
 \item{For direct computation of steady states and relaxation dynamics;
 this may be useful for complex systems because of the simplicity of the
 algorithm and resulting formulas and because often we do not know the  rate
 constants for complex networks, and kinetics that is ruled by orderings
 rather than by exact values of rate constants may be very useful in
 practically frequent situation when the values of the various reaction constants
 are unknown or poorly known;}
 \item{For planning experiments and mining the experimental data
 -- the observable kinetics is more sensitive to reactions from
 the dominant network, and much less sensitive to other reactions,
 the relaxation spectrum of the dominant network is  explicitly connected
 with the correspondent reaction rate constants,
 and the  eigenvectors  (``modes") are sensitive to the constant ordering,
 but not to exact values;}
 \item{The steady states and dynamics of the dominant system could
 serve as a robust first approximation in perturbation theory or as a
 preconditioning in numerical methods.}
\end{itemize}

The next step should be development of asymptotic estimates for
networks with modular structure and time separations between
modules, not between individual reactions. But now it seems that the
most important further development should be the asymptotology of
nonlinear reaction networks. For multiscale nonlinear reaction
networks the expected dynamical behaviour is to be approximated by
the system of dominant networks. These networks may change in time
(this is the significant difference from the linear case) but remain
relatively simple.

\section*{Appendix: Mathematical Backgrounds of Accuracy Estimation}

\subsection*{Estimates for Perturbed Acyclic Networks}

The famous Gerschgorin theorem (\cite{MM}, \cite{Varga}) gives
estimates of eigenvalues. We need also estimates of eigenvectors.
Below $A=(a_{ij})$ is a complex $n\times n$ matrix, $Q_i = \sum_{j,
j\neq i} |a_{ji}|$ (sums of non-diagonal elements in columns).

{\bf Gerschgorin theorem} (\cite{MM}, p.~146): The characteristic
roots of $A$ lie in the closed region $G^Q$ of the $z$-plane
\begin{equation}
G^Q=\bigcup_i G^Q_i \; \; (G^Q_i=\{z \, \bigl| \, |z-a_{ii}|\leq
Q_i\}.
\end{equation}
Areas $G^Q_i$ are the Gerschgorin discs. (The same estimate are
valid for sums in rows, $P_i$. Here and below we don't duplicate the
estimates.)

Gerschgorin disks $G^Q_i$ ($i=1, \ldots n$) are isolated, if $G^Q_i
\cap G^Q_j= \varnothing$ for $i\neq j$. If disks $G^P_i$ ($i=1,
\ldots n$) are isolated, then the spectrum of $A$ is simple, and
each Gerschgorin disk $G^Q_i$ contains one and only one eigenvalue
of $A$ (\cite{MM}, p.~147).

We assume that Gerschgorin disks $G^Q_i$ ($i=1, \ldots n$) are
isolated: for all $i,j$ ($i\neq j$)
\begin{equation}
|a_{ii}-a_{jj}| > Q_i + Q_j.
\end{equation}
Let us introduce the following notations:
\begin{equation}
\begin{split}
&\frac{Q_i}{|a_{ii}|}=\varepsilon_i, \; \varepsilon=\max_i
\varepsilon_i, \; \frac{|a_{ij}|}{|a_{jj}|} = \chi_{ij}, \;
\chi=\max_{i,j, i\neq j} \chi_{ij},  \\ & g_i=\min_{j, j\neq i}
\frac{|a_{ii}-a_{jj}|}{|a_{ii}|}, \;  g=\min_i g_i.
\end{split}
\end{equation}
Usually, we consider $\varepsilon_i$ and $\chi_{ij}$ as sufficiently
small numbers. In contrary, the {\it diagonal gap} $g$ should not be
small, (this is the {\it gap condition}). For example, if for any
two diagonal elements $a_{ii}$, $a_{jj}$ either $a_{ii} \gg a_{jj}$
or $a_{ii} \ll a_{jj}$, then $g_i \gtrsim 1$ for all $i$.

Let $\lambda_i \in G^Q_i$ be the eigenvalue of $A$ ($|\lambda_i  -
a_{11}| < Q_1$). Let us estimate the corresponding right eigenvector
$r^{(i)}$. We take $r^i_i=1$ and for $j\neq i$ introduce a
$(n-1)$-dimensional vector $\tilde{x}^i$: $\tilde{x}^i_j = r^i_j
(a_{jj}-a_{ii})$ ($i\neq j$). For $\tilde{x}^i$ we get equation
\begin{equation}\label{eigenequat}
(1-B^{(i)} )\tilde{x}^i =-\tilde{a}^i
\end{equation}
where $\tilde{a}^i$ is a vector of the non-diagonal elements of the
$i$th column of $A$ ($\tilde{a}^i_j=a_{ij}$, $j\neq i$), and the
$(n-1) \times (n-1)$ matrix $B^i$ has matrix elements ($j,l \neq i$)
\begin{equation}\label{matrixB}
b^{(i)}_{jj}=\frac{\lambda_i-a_{ii}}{a_{jj}-a_{ii}}, \;\;
b^{(i)}_{jl}=\frac{a_{jl}}{a_{ll}-a_{ii}} \; (l \neq j)
\end{equation}
Due to the Gerschgorin estimate,
$|b^{(i)}_{jj}|<\frac{Q_i}{|a_{jj}-a_{ii}|}$. From Eq.
(\ref{eigenequat}) we obtain:
\begin{equation}\label{Correction}
\tilde{x}^i=-\tilde{a}^i -B^{(i)}(1-B^{(i)})^{-1}\tilde{a}^i .
\end{equation}
From this definition and simple estimates in $l^1$ norm, we get the
following estimate of eigenvectors.

{\bf Theorem 2}. Let the Gerschgoring disks be isolated, and the
diagonal gap be big enough: $g>n \varepsilon$. Then for the $i$th
eigenvector of $A$ the following uniform estimate holds:
\begin{equation}\label{firOrEs}
|r^i_j| \leq \frac{\chi}{g}+\frac{n \varepsilon^2}{g(g-n \varepsilon
)}  \; \; (j\neq 1, \ r^i_i=1). \; \; \square
\end{equation}

So, if the matrix  $A$ is diagonally dominant and the diagonal gap
$g$ is big enough, then the eigenvectors are proven to be close to
the standard basis vectors with explicit evaluation of accuracy.

The first correction to eigenvectors is also given by Eq.
(\ref{Correction}). If for the iteration we use the Gerschgorin
estimates for eigenvalue $\lambda_i\approx a_{ii}$, then we can
write in the next approximation for eigenvectors ($r^i_i=1, j\neq
i$):
\begin{equation}\label{eigenvectorcorrection}
r^i_j=- \frac{a_{ji}}{a_{jj}-a_{ii}}-\frac{(B^{(i)}_{\rm nd
}(1-B^{(i)}_{\rm nd })^{-1}\tilde{a}^i)_j}{a_{jj}-a_{ii}}
\end{equation}
where $B^{(i)}_{\rm nd }$ is the non-diagonal part of $B^{(i)}$: it
has the same non-diagonal elements and zeros on diagonal. There
exists plenty of further simplifications for this iteration formula.
For example, one can leave just the first term, that gives the first
order approximation in the power of $\varepsilon$ ($\chi \leq
\varepsilon$).

To apply these estimates to an acyclic network supplemented by
additional reactions, we have to use the eigenbasis of this acyclic
network (Sec.~\ref{sec4}). Direct use of this theorem and estimates
for a kinetic matrix $K$ in the standard basis is impossible, the
diagonal dominance in this coordinate system is not large, and sums
of elements in columns are zero. To apply this theorem we need two
lemmas.

Let $\mathcal{W}$ be a reaction network without branching (a finite
dynamical system) with $n$ vertices. Then the number of reactions in
$\mathcal{W}$ is $n-f$, where $f$ is the number of fixed points (the
vertices without outgoing reactions). Let $\Gamma$ be the set of
stoichiometric vectors for $\mathcal{W}$.

{\bf Lemma 1}. $\Gamma$ forms a basis in the subspace $\{c\, | \,
\sum_i c_i = 0\}$ if and only if the reaction network $\mathcal{W}$
is acyclic and connected (has only one fixed point). $\square$

Let us consider a general reaction network on the set $A_1, ...
A_n$. For stoichiometric vector of reaction $A_{i} \to A_l$ we use
notation $\gamma_{li}$. Assume that the auxiliary dynamical system
$i \mapsto \phi(i)$ for a given reaction network is acyclic and has
only one attractor, a fixed point. For this auxiliary network, we
use notation: $\kappa_i = k_{ji}$ for the only reaction $A_i \to
A_j$, or $\kappa_i = 0$.

For every reaction of the initial network, $A_{i} \to A_l$, a linear
operators $Q_{i l}$ can be defined by its action on the basis
vectors, $\gamma_{\phi(i)\, i}$:
\begin{equation}\label{OprQ}
Q_{i l} (\gamma_{\phi(i)\, i}) = \gamma_{li}, \; Q_{i l}(
\gamma_{\phi(p)\, p}) = 0 \; \mbox{for} \; p \neq i.
\end{equation}

{\bf Lemma 2}. The kinetic equation for the whole reaction network
(\ref{kinur}) could be transformed to the form
\begin{equation}\label{kinurWholeAcy}
\begin{split}
\frac{\D c}{\D t}&=\sum_i \left(1+ \sum_{l, \, l \neq \phi(i)}
\frac{k_{li}}{\kappa_i} Q_{il}\right) \gamma_{\phi(i)\, i} \kappa_i
c_i\\  &= \left(1+ \ \sum_{j,l \, (l \neq \phi(j))}
\frac{k_{lj}}{\kappa_j} Q_{jl}\right) \sum_i  \gamma_{\phi(i)\, i}
\kappa_i c_i \\ &=\left(1+ \ \sum_{j,l \, (l \neq \phi(j))}
\frac{k_{lj}}{\kappa_j} Q_{jl}\right) \tilde{K}c,
\end{split}
\end{equation}
where $\tilde{K}$ is kinetic matrix of the kinetic equation for the
auxiliary network. $\square$

By construction of auxiliary dynamical system, ${k_{li}} <
{\kappa_i}$ if ${l \neq \phi(i)}$, and for reaction networks with
well separated constants ${k_{li}} \ll {\kappa_i}$. Notice also that
the matrix $Q_{jl}$ does not depend on rate constants values.

For matrix $\tilde{K}$ we have the eigenbasis in explicit form. Let
us represent system (\ref{kinurWholeAcy}) in this eigenbasis of
$\tilde{K}$. Any matrix $B$ in this eigenbasis has the form
$B=(\tilde{b}_{ij})$, $\tilde{b}_{ij}= l^i B r^j= \sum_{qs}l^i_q
b_{qs} r^j_s $, where $(b_{qs})$ is matrix $B$ in the initial basis,
$l^i$ and $r^j$ are left and right eigenvectors of $\tilde{K}$
(\ref{rightAcyc}), (\ref{leftAcyc}). In eigenbasis of $\tilde{K}$
the estimates of eigenvalues and estimates of eigenvectors are much
more efficient than in original coordinates: the system is strongly
diagonally dominant. Transformation to this basis is an effective
preconditioning for the perturbation theory that uses auxiliary
kinetics as a first approximation to the kinetics of the whole
system.

\subsection*{Estimates for Perturbed Ergodic Systems}

Let us consider a strongly connected network with kinetic matrix
$K$.  The corresponding kinetics is ergodic and there exists unique
normalized steady state $c_i^*>0$, $\sum_i c_i^*=1$. For each $i$ we
define $\kappa_i = \sum_j k_{ji}$. The number $-\kappa_i$ is the
$ii$th diagonal element of unperturbed kinetic matrix $K$.

Let this network be perturbed by outgoing reactions $A_i \to 0$. The
perturbation has the ``loss form": the perturbed matrix is $K-{\rm
diag}(\varepsilon_i \kappa_i)$, perturbation of each diagonal
element is relatively small (diag is the diagonal matrix).

The perturbations $\varepsilon_i \kappa_i$ are {\it relatively
small} with respect to $\kappa_i$, but not obligatory small with
respect to other rate constants.

First, we do not assume anything about value of $\varepsilon_i \geq
0$ and make the following transformation. For an arbitrary
normalized vector $r$ ($r_i \geq 0$, $\sum_i r_i =1$) we add to the
network reactions $A_i \to A_j$ with reaction rates $q_{ji}=r_j
\varepsilon_i \kappa_i$. We use  $Q(r)$ for the kinetic matrix of
this additional network. Simple algebra gives
\begin{equation}
\begin{split}
Q(r)+{\rm diag}(\varepsilon_i \kappa_i)&=
 [\varepsilon_1 \kappa_1 r, \varepsilon_2 \kappa_2 r,... \varepsilon_n
 \kappa_n r] \\
 &=r (\varepsilon_1 \kappa_1, \varepsilon_2 \kappa_2,... \varepsilon_n
 \kappa_n).
\end{split}
\end{equation}
Here, in the right hand side we have a matrix, all columns of which
are proportional to the vector $r$, this is a product of $r$ on the
vector-raw of coefficients. We represent the perturbed matrix in the
form $K-{\rm diag}(\varepsilon_i \kappa_i) = K+Q(r)-(Q(r)+{\rm
diag}(\varepsilon_i \kappa_i))$.

{\bf Theorem 3.} There exists such normalized positive $r^*$ that
$(K+Q(r^*))r^*=0$. This $r^*$ is an eigenvector of the perturbed
network with the eigenvalue $\lambda = \sum_i r^*_i \varepsilon_i
\kappa_i$, and, at the same time, it is a steady-state for the
network with kinetic matrix $K+Q(r^*)$.

To prove existence it is sufficient to mention, that for any $r$ the
network with kinetic matrix $K+Q(r)$ has unique positive normalized
steady state $c^*(r)$, which depends continuously on $r$. The map $r
\mapsto c^*(r)$ has a fixed point $r^*$ (the Brouwer fixed point
theorem). $\square$

This representation allows us to produce useful estimates, for
example, when the unperturbed system is a cycle,  we find
$|r^*_i-c^*_i| < 3 \varepsilon |c^*_i|$ under condition
$\varepsilon< 0.25$, where $\varepsilon =\sum \varepsilon_i$.
Formula for the first correction gives ($r^*=c^*_i+\delta r_i$,
$w=k_i c^*_i$):
\begin{equation}\label{circlecorrection}
\begin{split}
\delta r_i = \frac{v_i}{k_i}, \; v_i = v+w \sum_{j=1}^i (\varepsilon
c^*_j - \varepsilon_j), \\ v= \frac{w}{n} \sum_{i=1}^n i(\varepsilon
c^*_i - \varepsilon_i).
\end{split}
\end{equation}
For more complex networks, the explicit formulas for corrections
could be produced on the base of the network graphs, similar to the
steady-state formulas, presented, for example, by \cite{Yab}.

So, the asymptotic analysis gives good approximation of eigenvectors
and eigenvalues for kinetic matrix. The condition number is big
(unbounded) but these estimates work even better when the constants
become more separated. Nevertheless, some caution is needed: the
error is proven to be small, but the residuals (the values
$\|Kr-\lambda r\|$ for approximations of $r$ and $\lambda$) may be
not small (\cite{GorbaRadul2008}).


\begin{thebibliography}{1}
{\small

\bibitem[Andrianov \& Manevitch(2002)]{AndrianovManevitch2002}Andrianov, I.V. \& Manevitch, L.I. (2002). Asymptotology: Ideas,
Methods and Applications (Series: Mathematics and Its Applications,
Vol. 551), Dordrecht--Boston--London:  Springer.

\bibitem[Antoulas \& Sorensen(2002)]{AntoulasSorensen2002}Antoulas, A.C. \& Sorensen, D.C. (2002). The Sylvester
equation and approximate balanced reduction, {\em Linear Algebra and
Its Applications, 351-352}, 671--700.

\bibitem[Aris(1965)]{Aris1965}Aris, R. (1965). {\em Introduction to the Analysis of Chemical Reactors},
Englewood Cliffs, New Jersey: Prentice-Hall, Inc.

\bibitem[Balian, Alhassid \& Reinhardt(1986)]{Bal}Balian, R., Alhassid, Y., \& Reinhardt, H. (1986). Dissipation in many--body systems:
A geometric approach based on information theory,  {\em Physics
Reports  131 (1 )}, 1--146.

\bibitem[Bodenstein(1913)]{Bodenstein1913}Bodenstein, M. (1913). Eine Theorie der Photochemischen
Reaktionsgeschwindigkeiten, {\em Z . Phys. Chem.  85},  329--397.

\bibitem[Boyd(1978)]{Boyd}Boyd, R.~K. (1978). Some common oversimplifications in teaching chemical
kinetics, {\em J. Chem. Educ. 55,} 84--89.

\bibitem[Brown \& Cooper(1993)]{BrownCo}Brown, G.~C., \& Cooper, C.~E. (1993). Control analysis applied
to a single enzymes: can an isolated enzyme have a unique
rate--limiting step? {\em Biochem. J. 294,} 87--94.

\bibitem[Bykov, Goldfarb, Gol'dshtein, \& Maas, U.
(2006)]{BGGMaas2006} Bykov, V., Goldfarb, I., Gol'dshtein, V.,
Maas, U. (2006). On a Modified Version of ILDM Approach:
Asymptotical Analysis Based on Integral Manifolds Method, {\it IMA
J. of Applied Mathematics 71 (3)}, 359--382.

\bibitem[Christiansen(1953)]{Christiansen1953}Christiansen, J.A. (1953). The Elucidation of Reaction Mechanisms by
the Method of Intermediates in Quasi-Stationary Concentrations, {\em
Adv. Catal. 5}, 311--353.

\bibitem[Condon \& Ivanov(2004)]{CondonTruncation2004}Condon,
    M. \& Ivanov, R. (2004).  Empirical Balanced Truncation of
    Nonlinear Systems, {\em J. Nonlinear Sci. 14}, 405--414.

\bibitem[Cornish-Bowden \& Cardenas(1990)]{CorBow}Cornish-Bowden, A. \& Cardenas, M.~L. (1990). {\em Control on Metabolic
Processes}, New York: Plenum Press.

\bibitem[Coxson \& Bischoff(1987)]{LumpingOBservability}Coxson,
    P.G.\& Bischoff, K.B., (1987). Lumping strategy. 2. System
    theoretic approach,  {\em Ind. Eng. Chem. Res.,  26 (10)},
    2151--2157.

\bibitem[Djouad \& Sportisse(2002)]{NonstiffAtmospheric2002}Djouad, R. \& Sportisse, B. (2002). Partitioning techniques and
lumping computation for reducing chemical kinetics. APLA: An
automatic partitioning and lumping algorithm, {\em Applied
Numerical Mathematics, 43 (4)}, 383--398.

\bibitem[Dobrushin(1956)]{DobrushinErgCoeff1956}Dobrushin, R.L. (1956). Central limit theorem for non-stationary
Markov chains I, II, {\em Theor. Prob. Appl. 1}, 163--80, 329--383.

\bibitem[Dokoumetzidis \& Aarons(2009)]{PropLumpSysBio2009}Dokoumetzidis, A. \& Aarons, L.
(2009).
Proper lumping in systems biology models, {\em IET Systems
Biology,  3 (1)}, 40--51.

\bibitem[Farkas(1999)]{LumpingStructure}Farkas, G. (1999).
Kinetic lumping schemes, {\em Chem. Eng. Sci., 54 (17)},
3909--3915.

\bibitem[Feinberg(1972)]{Feinberg1972}Feinberg, M. (1972). On
chemical kinetics of a certain class, {\em Arch. Rat. Mech. Anal.
46 (1)}, 1--41.

\bibitem[Feng, Hooshangi, Chen, Li, Weiss, \& Rabitz(2004)]{Rabi3}Feng, X-J., Hooshangi, S., Chen, D.,
Li, G., Weiss, R., \& Rabitz, H. (2004). Optimizing Genetic Circuits
by Global Sensitivity Analysis, {\em Biophys J. 87}, 2195--2202.

\bibitem[Gibbs(1902)]{Gibb}Gibbs, G.W. (1902). {\em Elementary Principles in Statistical
Mechanics,} New Haven: Yale University Press.

\bibitem[Golub \& Van Loan(1996)]{Golub}Golub, G.H. \& Van Loan, C.F. (1996). Matrix Computations (3rd
edition), Baltimore: The Johns Hopkins University Press.

\bibitem[Gorban(1984)]{Gorban1984}Gorban, A.~N. (1984).
{\it Equilibrium encircling. Equations of chemical kinetics and their
thermodynamic analysis}, Nauka, Novosibirsk.

\bibitem[Gorban,  Bykov, \& Yablonskii(1986)]{without}Gorban, A.~N.,
Bykov, V.~I., \& Yablonskii, G.~S. (1986). Thermodynamic function
analogue for reactions proceeding without interaction of various
substances, {\em Chem. Eng. Sci. { 41} (11)}, 2739--2745.

\bibitem[Gorban, Bykov \& Yablonskii(1986)]{Ocherki}Gorban, A.~N.,  Bykov, V.~I., \& Yablonskii G.~S.
(1986). {\em Essays on chemical relaxation,} Novosibirsk: Nauka.

\bibitem[Gorban \& Karlin(2003)]{InChLANL}Gorban, A.~N., \& Karlin, I.~V. (2003). Method of invariant manifold for chemical
kinetics, {\em Chem. Eng. Sci.  58,} 4751--4768.

\bibitem[Gorban \& Karlin(2005)]{GorKar}
Gorban, A.~N., \& Karlin, I.~V. (2005).
\newblock {\em Invariant manifolds for physical and chemical kinetics}, volume
  660 of Lect. Notes Phys.
\newblock Berlin--Heidelberg--New York: Springer.

\bibitem[Gorban, Karlin, Ilg, \& \"{O}ttinger(2001)]{GKIOeNONNEWT2001}Gorban, A.N., Karlin,  I.V., Ilg, P.,
\& \"{O}ttinger, H.C. (2001). Corrections and enhancements of
quasi--equilibrium states, J.Non--Newtonian Fluid Mech. {\bf 96}
(2001), 203--219.

\bibitem[Gorban, Karlin, \& Zinovyev(2004)]{Grids}
Gorban, A.N., Karlin, I.V., Zinovyev, A.Yu. (2004). Invariant grids for reaction kinetics,
{\em Physica A { 333}} (2004), 106--154. Preprint online:
http://www.ihes.fr/PREPRINTS/P03/Resu/resu-P03--42.html

\bibitem[Gorban \& Radulescu(2008)]{GorbaRadul2008}Gorban, A.~N. \& Radulescu, O. (2008). Dynamic
and static limitation in reaction networks, revisited, {\em
Advances in Chemical Engineering 34}, {103-173}; e-print:
{http://arxiv.org/abs/physics/0703278}

\bibitem[Greuel \& Pfister(2002)]{orderPfi}Greuel, G.-M. \& Pfister, G. (2002). {\em A Singular Introduction to Commutative
Algebra,} Berlin--Heidelberg--New York:  Springer.

\bibitem[Gugercin \&
    Antoulas(2004)]{GugercinAntoulas2004}Gugercin, S. \&
    Antoulas, A.C. (2004). A survey of model reduction by
    balanced truncation and some new results, {\em Int. J.
    Control, 77 (8)}, 748--766.

\bibitem[Hangos, Bokor, \& Szederk\'enyi(2004)]{Hangos2004}Hangos, K.M.,
Bokor, J., \& Szederk\'enyi G. (2004). {\em Analysis and Control of
Nonlinear Process Systems}, London: Springer-Verlag.

\bibitem[Hangos \& Cameron(2001)]{HangosProcessBook}Hangos, K.M.
\& Cameron, I.T. (2001), Process Modelling and Model Analysis.
London: Academic Press.

\bibitem[Helfferich(1989)]{Helfferich1989}Helfferich, F.G. (1989). Systematic approach to elucidation of
multistep reaction networks, {\em J. Phys. Chem. 93 (18)},
6676--6681

\bibitem[Hutchinson \& Luss(1970)]{LumpParal1stOrd}Hutchinson, P. \& Luss, D. (1970), Lumping of mixtures with
many parallel first order reactions: Chemical Engineering journal,
1, 129--135.

\bibitem[Jaynes(1963)]{Janes1}Jaynes, E.T. (1963). Information theory and statistical mechanics, {\em in:
Statistical Physics. Brandeis Lectures, V.3, K. W. Ford, ed.}, New
York: Benjamin,  pp. 160--185.

\bibitem[Johnston(1966)]{Johnston}Johnston, H.~S. (1966).
{\em Gas phase reaction theory,}
New York: Roland Press.

\bibitem[Kazantzis \& Kravaris(2006)]{KazKraLya}
Kazantzis, N. \& Kravaris,  C. (2006). A New Model Reduction
Method for Nonlinear Dynamical Systems using Singular PDE Theory,
In: {\it Model Reduction and Coarse-Graining Approaches for
Multiscale Phenomena, A.N. Gorban, N. Kazantzis, Y.G. Kevrekidis,
H.C. Ottinger and C. Theodoropoulos (eds.),} Springer, 3--15.

\bibitem[Klonowski(1983)]{Klonowski1983}Klonowski, W. (1983).  Simplifying Principles for Chemical and
Enzyme Reaction Kinetics, {\em Biophys.Chem. 18}, 73--87.

\bibitem[Kruskal(1963)]{Kruskal}Kruskal, M.~D.  (1963).
Asymptotology, In: {\em Mathematical Models in Physical
Sciences,} ed. by S. Dobrot, Prentice-Hall, New Jersey, Englewood
Cliffs, 17--48.

\bibitem[Kuo \& Wei(1969)]{LumpWei2}Kuo, J.~C. \& Wei, J. (1969). A lumping analysis in monomolecular reaction
systems. Analysis of the approximately lumpable system. {\em Ind.
Eng. Chem. Fundam. 8,} 124--133.

\bibitem[Lam(1993)]{Lam1993}Lam, S.~H. (1993). Using CSP to
    Understand Complex Chemical Kinetics, {\em Combustion
    Science and Technology, 89 (5)}, 375--404.

\bibitem[Lall, Marsden \& Glavaki(2002)]{MarsdenTruncation2000}Lall, S., Marsden, J.E., \& Glavaki, S. (2002). A subspace approach to
balanced truncation for model reduction of nonlinear control
systems, {\em Int. J. Robust Nonlinear Control 12 (6)}, 519--535.

\bibitem[Lam \& Goussis(1994)]{LamGous1994}Lam, S.~H., \& Goussis, D.~A. (1994).
The CSP Method for Simplifying Kinetics, {\em International
Journal of Chemical Kinetics  26}, 461--486.

\bibitem[Li \& Rabitz(1989)]{LumpLiRab1}Li, G., \& Rabitz,  H. (1989).
A general analysis of exact lumping in chemical
kinetics. {\em Chem. Eng. Sci. 44,} 1413--1430.

\bibitem[Liao \& Lightfoot(1988)]{LumpLiaoBiochem}Liao, J.~C. \& Lightfoot~Jr. E.~N. (1988). Lumping analysis of biochemical
reaction systems with time scale separation, {\em Biotechnology
and Bioengineering 31,} 869--879.

\bibitem[Lidskii(1965)]{Lid}Lidskii, V. (1965). Perturbation theory of non-conjugate operators.
{\em U.S.S.R. Comput. Math. and Math. Phys., 6,} 73--85.

\bibitem[Lin, Leibovici \& Jorgensen(2008)]{OptimalLumping2008}Lin, B., Leibovici, C.F., Jorgensen, S.B. (2008), Optimal
component lumping: Problem formulation and solution techniques,
{\em Computers \& Chemical Engineering, 32,} 1167--1172.

\bibitem[Litvinov \& Maslov(2005)]{LitMas}Litvinov, G.~L. \& Maslov, V.~P. (Eds.) (2005). {\em Idempotent mathematics and
mathematical physics, Contemporary Mathematics,}  Providence: AMS.


\bibitem[Maas, \& Pope(1992)]{Maas}Maas, U., \& Pope, S.B.  (1992).
Simplifying chemical kinetics: intrinsic low -- dimensional
manifolds in composition space, {\it Combustion and Flame 88},
239--264.

\bibitem[Marcus \& Minc(1992)]{MM}Marcus, M. \& Minc,  H. (1992). {\em A survey of matrix theory and
matrix inequalities,} New-York: Dover.

\bibitem[Maria(2006)]{LumpingLiving2006}Maria, G. (2006),
Application of lumping analysis in modelling
the living systems : A trade-off between simplicity and model
quality, {\em Chemical and Biochemical Engineering Quarterly, 20
(4)}, 353--373.

\bibitem[Meyn(2007)]{MeynNets2007}Meyn, S.R. (2007). {\em Control Techniques for Complex Networks},
Cambridge University Press, Cambridge.

\bibitem[Meyn \& Tweedie(2009)]{MeynMarkCh2009}Meyn, S.P. \& Tweedie, R.L. (2009). {\em Markov Chains and
Stochastic Stability}, 2nd Edition, Cambridge: Cambridge University
Press.

\bibitem[Moore(1981)]{Moore1981}Moore, B.C. (1981) Principal component analysis in linear
system: controllability, observability and model reduction. {\em
IEEE Transactions on Automatic Control}, AC-26.

\bibitem[Northrop(1981)]{Northrop1}Northrop, D.~B. (1981). Minimal kinetic mechanism and general
equation for deiterium isotope effects on enzymic reactions:
uncertainty in detecting a rate-limiting step, {\em Biochemistry,
20,} 4056--4061.

\bibitem[Northrop(2001)]{Northrop2}Northrop, D.~B. (2001).
Uses of isotope effects in the study of
enzymes, {\em Methods 24,} 117--124.

\bibitem[Pepiot-Desjardins \& Pitsch(2008)]{LumpingCombust2008}Pepiot-Desjardins, P.,
Pitsch, H. (2008). An automatic chemical lumping method for the
reduction of large chemical kinetic mechanisms {\em Combustion
Theory and Modelling,  12 {6}}, 1089--1108.

\bibitem[Prigogine \& Defay(1954)]{PrigogineDefay1954}Prigogine, I. \& Defay, R. (1954).
{\em Chemical Thermodynamics} London: Longmans.

\bibitem[Procaccia \& Ross(1977)]{ProcacciaRoss}Procaccia, I. \& Ross, J. (1977). Stability and relative
stability in reactive systems far from equilibrium. I.
Thermodynamic analysis J. Chem. Phys. 67, 5558--5564.

\bibitem[Radulescu, Gorban, Zinovyev \& Lilienbaum(2008)]{RadGorZinLil2008}Radulescu, O., Gorban,
A., Zinovyev, A., \& Lilienbaum, A. (2008). Robust simplifications
of multiscale biochemical networks, {\em BMC Systems Biology 2
(1)}, 86 http://www.biomedcentral.com/1752-0509/2/86

\bibitem[Rate-controlling step(2007)]{R-cont}Rate-controlling step (2007).
In: IUPAC Compendium of Chemical Terminology,
E-version, http://goldbook.iupac.org/R05139.html.

\bibitem[Ray(1983)]{Ray}Ray, W.~J. (Jr.) (1983). A rate--limiting step: a quantitative
definition. Application to steady--state enzymic reactions, {\em
Biochemistry}, 22, 4625--4637.

\bibitem[Robbiano(1985)]{orderRobb}Robbiano, L. (1985). Term orderings on the polynomial ring, In: {\em Proc.
EUROCAL 85, vol. 2, ed. by B.~F.~Caviness, Lec. Notes in Computer
Sciences 204,}  Berlin--Heidelberg--New York--Tokyo: Springer,
513--518.

\bibitem[Roussel \& Fraser(1991)]{Roussel91}Roussel, M.R., \& Fraser, S.J. (1991).
On the geometry of transient relaxation. {\it J.\ Chem.\ Phys.
94}, 7106--7111.

\bibitem[Segel \& Slemrod(1989)]{Segel89}Segel, L.A., \& Slemrod, M. (1989).
The quasi-steady-state assumption: A case study in perturbation.
{\it SIAM Rev. {31}}, 446-477.

\bibitem[Semenov(1939)]{Semenov1939}Semenov, N.N. (1939). On the Kinetics of Complex Reactions, {\em J.
Chem. Phys. 7}, 683--699.

\bibitem[Seneta(1981)]{Seneta1981}Seneta, E. (1981). {\em Nonnegative Matrices and Markov Chains},
Springer, New York.

\bibitem[Stueckelberg(1952)]{Stueck}Stueckelberg, E.C.G. (1952). Theoreme $H$ et unitarite de
$S$, {\em Helv. Phys. Acta { 25} (5)}, 577--580.

\bibitem[Temkin, Zeigarnik, \& Bonchev(1996)]{Temkin1996}Temkin, O.N., Zeigarnik, A.V., \& Bonchev, D.G. (1996). {\em
Chemical Reaction Networks: A Graph-Theoretical Approach}, Boca
Raton, FL: CRC Press.

\bibitem[Toth, Li, Rabitz, \& Tomlin(1997)]{LumpLiRab2}Toth, J., Li, G., Rabitz, H., \& Tomlin, A.~S. (1997).
The Effect of Lumping and Expanding on Kinetic Differential
Equations, {\em SIAM J. Appl. Math. 57,} 1531--1556.

\bibitem[Turanyi, Tomlin, \&  Pilling(1993)]{Tomlin1993}Turanyi, T.,  Tomlin, A.S., \&  Pilling, M.J. (1993). On the error
of the quasi-steady-state approximation, {\em J. Phys. Chem.  97
(1)}, 163--172.

\bibitem[Van Mieghem(2006)]{VanMieghem2006}Van Mieghem, P. (2006). {\em Performance Analysis of
Communications Networks and Systems}, Cambridge University Press,
Cambridge.

\bibitem[Varga(2004)]{Varga}Varga, R.S. (2004). {\em Gerschgorin and His Circles, Springer series in
computational Mathematics, 36,}  Berlin -- Heidelberg -- New York:
Springer.

\bibitem[Vishik \& Ljusternik(1960)]{ViLju}Vishik, M.~I., \&
    Ljusternik, L.~A. (1960). Solution of some perturbation
    problems in the case of matrices and self-adjoint or
    non-selfadjoint differential equations. I, {\em Russian
    Math. Surveys, 15,} 1--73.

\bibitem[Vora \& Daoutidis(2001)]{Daoutidis}Vora, N., \& Daoutidis, P.
(2001). Nonlinear Model Reduction of Chemical Reaction Systems
{\em AIChE Journal, 47 (10)}, 2320--2332.

\bibitem[Wei \& Prater(1962)]{Wei62}Wei, J.,  \& Prater, C.  (1962).
The structure and analysis of complex reaction systems. {\it Adv.\
Catalysis, 13}, 203--393.

\bibitem[Wei \& Kuo(1969)]{LumpWei1}Wei, J., \& Kuo, J.~C. (1969).
A lumping analysis in monomolecular
reaction systems: Analysis of the exactly lumpable system, {\em
Ind. Eng. Chem. Fundam., 8,} 114--123.

\bibitem[White(2006)]{White}White, R.~B. (2006). {\it Asymptotic Analysis of Differential
Equations,} London: Imperial College Press \& World Scientific.

\bibitem[Whitehouse, Tomlin, \&
    Pilling(2004)]{LumpAthmTime}Whitehouse, L.~E., Tomlin,
    A.~S., \&  Pilling, M.~J. (2004). Systematic reduction of
    complex tropospheric chemical mechanisms, Part II: Lumping
    using a time-scale based approach, {\em Atmos. Chem. Phys.,
    4,} 2057--2081.

\bibitem[Yablonskii, Bykov, Gorban, \& Elokhin(1991)]{Yab}
Yablonskii, G.~S.,   Bykov, V.~I., Gorban, A.~N., \& Elokhin,
V.~I. (1991). {\em Kinetic models of catalytic reactions.
Comprehensive Chemical Kinetics, Vol. 32,} Compton R.~G. ed.,
Amsterdam: Elsevier.

\bibitem[Yablonsky, Mareels, \& Lazman(2003)]{YabCritSimpl}Yablonsky, G.S.,
Mareels, I.M.Y., Lazman, M. (2003). The Principle
of Critical Simplification in Chemical Kinetics, {\em Chem. Eng.
Sci. 58}, 4833--4842.

\bibitem[Yablonsky, Olea, \& Marin(2003)]{TAP}Yablonsky, G.S.,
Olea, M., \& Marin, G.B. (2003). Temporal Analysis of
Products (TAP): Basic Principles, Applications and Theory, {\em
Journal of Catalysis 216}, 120--134.

\bibitem[Zagaris, Kaper, \& Kaper(2004)]{ZaKapers}Zagaris, A.,
    Kaper, H.G., Kaper, T.J. (2004). Analysis of the
    computational singular perturbation reduction method for
    chemical kinetics, {\it J. Nonlinear Sci. 14,} 59--91.

\bibitem[Zavala \& Rodriguez \& Vargas-Villamil(2004)]{LumpingPetroleum2004}Zavala,
C.D., Rodriguez, J.E.R., Vargas-Villamil, F.D. (2004), An
algorithm for pseudocompound delumping and lumping into homologous
groups, {\em Petroleum Science and Technology, 22 (1-2)}, 45--60.


}

\end{thebibliography}
\end{document}